\documentclass{article}

\usepackage{arxiv}

\usepackage{amssymb}

\usepackage{tabularx}
\usepackage{graphicx}
\usepackage[export]{adjustbox}
\usepackage{subcaption}
\usepackage{mwe}
\usepackage{hyperref}
\usepackage[dvipsnames]{xcolor}
\usepackage{amsmath}
\usepackage{booktabs}
\usepackage{cuted}
\usepackage{comment}
\usepackage[english]{babel}
\usepackage{amsthm}
\usepackage{svg}

\usepackage{booktabs}
\usepackage{mdframed}

\newmdtheoremenv{defin}{Theorem}

\theoremstyle{definition}
\newtheorem{definition}{Definition}[section]

\theoremstyle{definition}

\theoremstyle{remark}

\title{Anomaly detection in cross-country money transfer temporal networks}

\author{
Salvatore Vilella \\
  DISIT,
  Universit\`a degli Studi del\\ Piemonte Orientale ``A. Avogadro''\\
  Viale Teresa Michel, 11\\
  Alessandria, Italy.  15121\\
  \texttt{{name}.{surname}@uniupo.it} \\
  \And
Arthur Thomas Edward Capozzi Lupi\\
  Dipartimento di Informatica\\
  Universit\`a degli Studi di Torino\\
  Corso Svizzera, 185\\
  Torino, Italy. 10149\\
  \texttt{{name}.{surname}@unito.it} \\
  \And
Marco Fornasiero,
Dario Moncalvo, 
Valeria Ricci,
Silvia Ronchiadin \\
  Anti Financial Crime Digital Hub\\
  Corso Inghilterra, 3\\
  Turin, Italy. 10138\\
  \texttt{{name}.{surname}@intesasanpaolo.com} \\
  \And
Giancarlo Ruffo \\
  DISIT,
  Universit\`a degli Studi del\\ Piemonte Orientale ``A. Avogadro''\\
  Viale Teresa Michel, 11\\
  Alessandria, Italy.  15121\\
  \texttt{{name}.{surname}@uniupo.it} \\
}

\begin{document}

\maketitle

\begin{abstract}This paper explores anomaly detection through temporal network analysis. Unlike many conventional methods, relying on rule-based algorithms or general machine learning approaches, our methodology leverages the evolving structure and relationships within temporal networks, that can be used to model financial transactions. Focusing on minimal changes in stable ecosystems, such as those found in large international financial institutions, our approach utilizes network centrality measures to gain insights into individual nodes. By monitoring the temporal evolution of centrality-based node rankings, our method effectively identifies abrupt shifts in the roles of specific nodes, prompting further investigation by domain experts.

To demonstrate its efficacy, our methodology is applied in the Anti-Financial Crime (AFC) domain, analyzing a substantial financial dataset comprising over 80 million cross-country wire transfers. The goal is to pinpoint outliers potentially involved in malicious activities, aligning with financial regulations. This approach serves as an initial stride towards automating AFC and Anti-Money Laundering (AML) processes, providing AFC officers with a comprehensive top-down view to enhance their efforts. It overcomes many limitations of current prevalent paradigms, offering a holistic interpretation of the financial data landscape and addressing potential blindness to phenomena that cannot be effectively estimated through single-node or narrowly focused transactional approaches.
\end{abstract}


\section{Introduction}\label{sec:intro}

Modern complex systems, encompassing fields as diverse as biology, cybersecurity, finance, healthcare,
and industrial processes, present an environment where deviations from the norm can hold critical significance. These deviations constitute a focal point for anomaly detection: they can signify a range of events depending on the domain, from subtle inefficiencies to outright malicious activities. 
Anomaly detection serves as a critical line of defense against system failures, security breaches or other potential threats. Early detection can mitigate the cascading effects that anomalies might trigger. By identifying deviations from the norm, anomaly detection allows to respond proactively, maintaining system integrity, data security, and operational efficiency; an ex-post identification of anomalies instead provides the analysts with useful insights on the nature of the system and its components, possibly informing them about potential future threats.

This is particularly true in the financial domain, where anomaly detection is a key tool for combating financial crime and for identifying suspicious or illegal activities. Depending on the specific tasks and needs of the controllers, both real-time approaches or post-event detection of anomalies can be exploited to unveil malicious activities. Currently, typical Anti Financial Crime (AFC) practices are rule-based~\footnote{According to the Correspondent Banking Due Diligence Questionnaire (CBDDQ) Guidance (Wolfsberg Group), https://t.ly/PIZms, last accessed: 17/08/2023.}. An important part in acquiring the necessary knowledge is played by surveys such as the Wolfsberg Group's CBDDQ and FCCQ, i.e., a long set of questions that are used by banks or other financial institutions to provide high-level information about their Financial Crime Compliance Programme; this and other AFC strategies, that are peculiar to each financial institution due to its own risk appetite and risk tolerance, are built upon complex layers of national and international regulations. The traditional approach of Anti Money Laundering toward data analysis is largely based on individual subjects as unit of analysis, in order to spot standalone customer behaviours that may suggest the presence of criminal activities leveraging a Financial Institution’s accounts and means of payment. Nevertheless, there is an important conceptual overlap between the traditional approaches to AFC and the analysis of complex systems: network analysis is an embedded activity in the investigation process leading to SAR (Suspicious Activity Report) already performed by human based reasoning, with the general support of individual productivity desktop tools, or of specifically developed automated systems. This activity, starting from previously selected starting points (suspicious activities already spotted), partially balance or complement the narrow focus coming from the customer-centric transaction monitoring output.

For these reasons, improving the state of the art on network analysis applied to the detection of anomalies in the financial domain could help regulators and financial institution in their fight against financial crimes, by switching from an \textit{atomic} to a \textit{high level} and \textit{context-aware} view of the system.
Nonetheless, detecting anomalies within complex systems poses a unique set of challenges. Traditional rule-based methods often fall short in capturing the subtle dynamics that characterize anomalies; on the other hand, machine learning and, more specifically, deep-learning based methods, while being very effective in capturing the embedded complexity of the system, can lack the \textit{interpretability} that regulators that regulators expect and law enforcement agencies need in order to take action. Furthermore, the domain experts are considered liable for not having reported a suspicious transaction to an extent that is proportional to the severity of the implausibility of the unreported case, thus calling for an anomaly detection system.

We position our work in this context: we propose a novel methodology, that exploits well consolidated network centrality measures and node rankings to unveil sudden and unexpected changes in the role of nodes in the system. This methodology aims to be applied to uncover potential anomalies in any system that can be modeled through network analysis. When applied to the financial domain, it will allow an AFC analyst to investigate for potential anomalies taking into account the activity of the actors and of their neighbours; this will be done in a fully interpretable way, relying on the clear meaning of the chosen centrality metrics. We will treat this as a problem of ranked information retrieval: we want our final output to be a list of observations, sorted by their measured deviation from the norm, so that outliers are prioritised in such a way that the probability of finding real anomalies towards the top of this list will be maximised.

\subsection{Assumptions and Requirements}\label{subsec:scope}
In this work we aim at developing a methodology based on temporal network analysis to detect anomalies on \textit{unlabeled} data, i.e., when no a-priori notion of \textit{anomaly} is provided. While our primary motivation was to elaborate a solution for the AFC domain, we believe that our approach can be extended to any context where the following assumptions are applicable:
\begin{itemize}
    \item \textbf{A1:} The \textit{domain experts}, in charge of overseeing the anomaly detection process, are requested to report a list of suspicious observations to a \textit{competent authority}.
    \item \textbf{A2:} Reports can only be submitted if the expert can provide a \textit{justification} for each detected anomaly, which could include a detailed explanation of the underlying dynamics leading to the identification, or also a human understandable interpretation of the algorithm's outcome.
    \item \textbf{A3:} The process of detecting and reporting anomalies is inherently \textit{human-in-the-loop} and, consequently, \textit{time-consuming}.
    \item \textbf{A4:} The human expert is \textit{not} supposed to receive a feedback from the competent authority confirming whether the reported anomalies correspond to actual violations; in the best scenario, the feedback provided to the human experts is partial and heavily delayed.
\end{itemize}

These assumptions help us identifying the following requirements that an ideal solution should have:

\begin{itemize}
    \item \textbf{R1:} It is not feasible to rely on a ground truth, since the competent authority will not necessarily validate the results; nevertheless, we can access to the list of anomalies considered \textit{relevant} for reporting purposes by the organization's domain expert.
    \item \textbf{R2:} Algorithms should provide experts with a concise list of potentially relevant outliers, ensuring the timely execution of their analysis and subsequent reporting to the competent authority.
    \item \textbf{R3:} The utilization of AI black boxes should be minimized as much as possible, unless their output are fully explainable.
\end{itemize}

Given the above assumptions and requirements, we will treat the task of anomaly detection as a problem of ranked information retrieval, namely by designing a process whose ultimate output is a ranking of potential anomalies, i.e., a list of observations sorted by their quantified deviations from a given norm. As argued, this should be done by always keeping the \textit{interpretability} of the results as a top priority. Therefore, we will define a detection pipeline that, always considering the above assumptions and requirements, will have as an output a set of nodes that, depending on the context of application, can be further inspected by a human expert. Indeed, our ultimate goal is neither to fully automate the process of anomaly detection, nor to predict anomalies, but rather to help the human eye in filtering the noise and in obtaining new insights that would potentially go undetected by following traditional practices. 

\begin{figure}[h!]
\centering
\includegraphics[width=\textwidth]{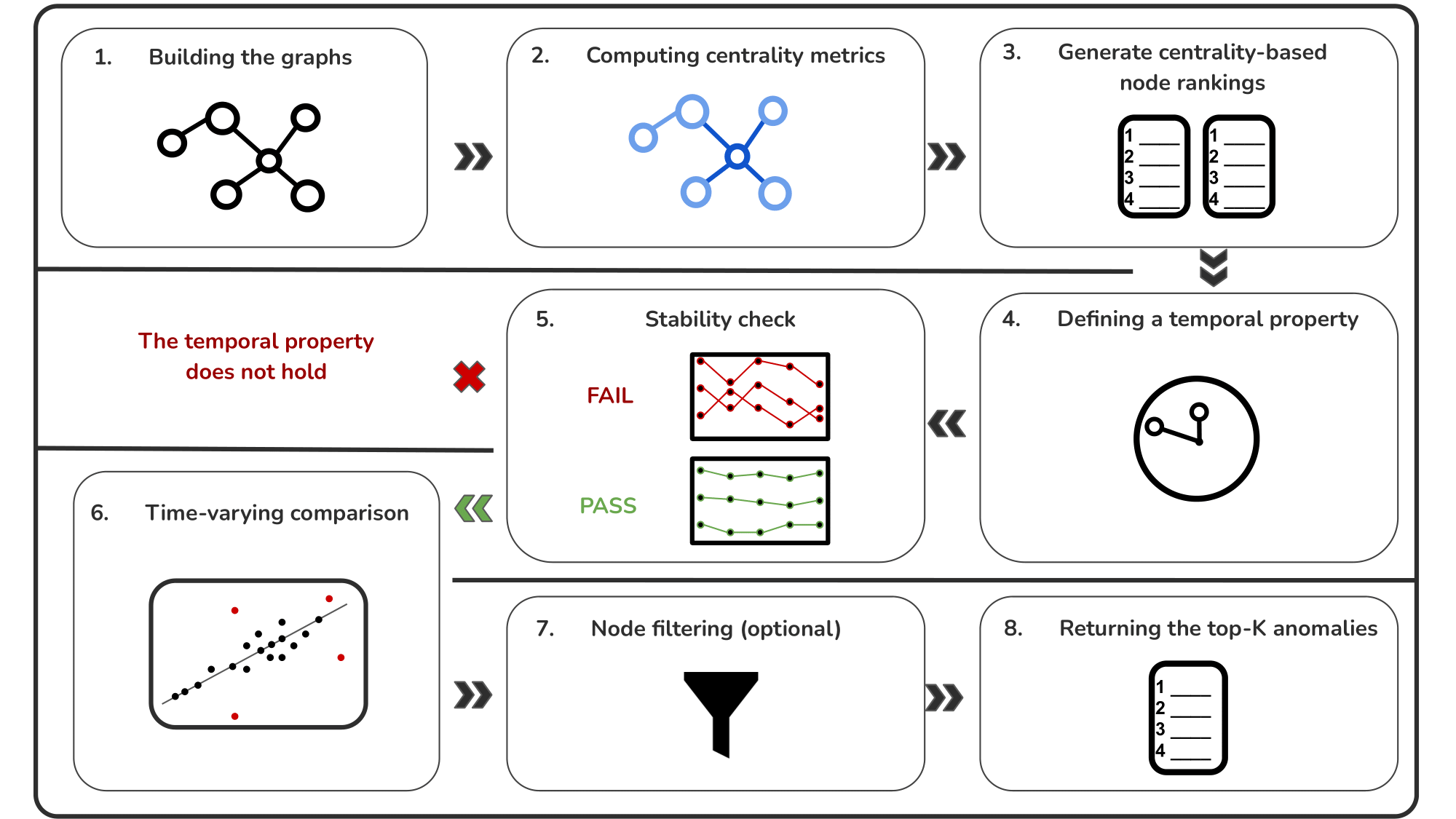} 
\caption{Scheme of the proposed pipeline of anomaly detection. Starting from raw transaction data, one or more final lists of potentially anomalous nodes will be yield as a final output. This is achieved by building financial graphs over time, keeping track of the change of network centrality of the nodes. This algorithm can also embed external domain knowledge, to facilitate the filtering of noise and let true anomalies emerge.}
\label{fig:scheme}
\end{figure}

While the details of the proposed methodology are provided in Sec.~\ref{sec:methods} and the pipeline, schematically shown in Fig.~\ref{fig:scheme}, is defined in Sec.~\ref{sec:proposal}, in Sec.~\ref{sec:casestudy}, we showcase its application to a very large, unlabeled, real-world dataset of cross-country wire transfers, adequately anonymized, provided by the bank Intesa Sanpaolo and the AFC Digital Hub (see the acknowledgements section at the end of the paper for more information about the consortium). We will show how our approach is capable of providing useful insights about the system and can also highlight individual anomalous nodes that are considered worth of further investigation by the domain experts.

\section{Related work}

\subsection{Anomaly detection on graphs}

Broadly speaking, there are many studies that focus on the problem of anomaly detection on networks.
Savage et al.~\cite{SAVAGE201462} review a number of relevant literature, observing that anomalies in online social networks (OSNs) can signify irregular, and often illegal behaviour. Detection of such anomalies has been used to identify malicious individuals, including spammers, sexual predators, and online fraudsters. In their survey, they list existing computational techniques for detecting anomalies in online social networks. Also, they characterize anomalies as being either static or dynamic, and as being labelled or unlabelled, and they survey methods for detecting these different types of anomalies. They also suggest that the detection of anomalies in online social networks is composed of two sub-processes; the selection and calculation of network features, and the classification of observations from this feature space. In addition, this survey provides an overview of the types of problems that anomaly detection can address and identifies key areas for future research.

A survey~\cite{KAUR2016199} also observes that anomalous activities in OSNs can represent unusual and illegal activities exhibiting different behaviors than others present in the same structure. Kaur et al. describe different types of anomalies and their novel categorization based on various characteristics. A review of a number of techniques for preventing and detecting anomalies along with underlying assumptions and reasons for the presence of such anomalies is covered as well, altogether with  data mining approaches used to detect anomalies. A special reference is made to the analysis of social network centred anomaly detection techniques which are broadly classified as behavior based, structure based and spectral based. Each one of these classifications further incorporates a wide number of techniques. In general, it is extremely common to use network analysis (NA) to extract network metrics that will inform different learning algorithms~\cite{van2015afraid,van2015guilt,van2017gotcha} also in the field of anomaly detection in transaction networks~\cite{van2015apate}.

Finally, Wang et al.~\cite{wang2017computing} address the problem of time-sensitive anomaly detection on road networks, in order to reveal unexpected traffic patterns whose identification could be helpful in road planning and management. They examine diverse anomalous traffic patterns, specifically identifying edges that exhibit complete disappearance or emergence between consecutive timeframes. They also consider patterns with a disparity in probabilities between the two time intervals exceeding a predefined threshold $\mu$. Interestingly, and similarly to our approach, they will rank the entries of their final output, evaluating the effectiveness of the anomaly detection in terms of precision-at-k.

\subsection{Anomaly detection in financial networks}
Many efforts have been put in the scientific literature to address the many challenges posed by anomaly detection on financial data. Examining the AFC problem through the lenses of computer science and complex systems allowed the development of many interesting approaches. Such approaches usually exploit machine learning, deep learning and complex networks - or a mixture of them - to tackle the problem of detecting anomalies in financial transactions. 

Indeed, complex networks are a powerful tool to model financial transactions: since financial data usually encodes an exchange of money between two or more entities, this interaction can easily be modeled as \textit{edges} between two or more \textit{nodes}. Network analysis (NA) applied to AFC, Anti-Money Laundering (AML), and more generally to the identification of anomalies in transaction networks is indeed growing in interest in recent years. García et al.~\cite{García2021} describe the research lines and developments based on NA carried out from 2015 to 2020 by the Spanish Tax Agency for tax control. They present case studies demonstrating how NA has been effectively applied in a real world scenario. On the one hand, pattern detection algorithms on graphs can facilitate the identification of frauds; on the other, community identification and community detection techniques help to provide a more precise picture of the economic reality.

Colladon and Remondi~\cite{FRONZETTICOLLADON201749} emphasize the role of social network metrics in the detection of money laundering practices related to companies. They map 4 different kinds of relational graphs to take into consideration different risk factors: economic sectors, geographical areas, amount of transactions, and links between different companies sharing the same owner or representatives. By means of a visual analysis of the fourth graph, the authors are able to identify clusters of subjects that were involved in court trials.

Garcia-Bedoya et al.~\cite{Bedoya2020} propose an approach based on AI and network analysis methodologies, underlining that many previous AML approaches fail to identify money laundering because they are based on static analysis after several days or months the financial transactions have been made. Furthermore, they point out that in a transaction network, two nodes involved in money laundering will have complex connections to mask their activities. In particular, they distinguish between 3 different categories of interactions that can hide an money laundering attempt:
\begin{enumerate}
    \item \textit{Path}: Node x uses intermediaries to send money to node y. Instead of a single transaction between x and y, we will have a path of transactions between the two nodes.
    \item \textit{Cycle}: The money starts from node x and after a path of transactions returns to the same node.
    \item \textit{Smurf}: Node x divides the transaction into many small transactions which, through intermediaries who can be both legal and natural persons, will arrive at node y.
\end{enumerate}
Liu et al.~\cite{Liu_Zhou_Zhu_Gu_He_2020}  argue that cycles formed with time-ordered transactions might be red flags in online transaction networks; a comprehensive study on the detection of motifs in financial graphs is carried in~\cite{jiang2022analyzing}, where Jiang explore motif-based embedding and its applications in a range of downstream graph analytical tasks. Starnini et al.~\cite{starnini2021smurf} also focus on smurfing, defining a method that efficiently finds suspicious smurf-like subgraphs.

Finally, also in~\cite{noble2003graph} unusual graph substructures are investigated. Here anomalous subgraph detection is also carried out by partitioning the graph into distinct subgraphs, each of which is tested against the others for unusual patterns. Noble at al. evaluate the graph regularity by means of conditional substructure entropy, a measure that defines the number of bits needed to describe an arbitrary substructure’s surroundings. 

Many studies described above are based on the identification of specific patterns and on the study of their temporal evolution. These approaches have the limitation of having to know a priori the patterns to look for within the transaction network. Alternative techniques are based on machine learning~\cite{Chen2018} or deep learning via graph neural networks (extensively reviewed in this survey~\cite{motie2023financial}), that generate compact vector representations for each node. Several approaches have been proposed and applied to fraud detection~\cite{9446887, dou2020enhancing, shi2022h2,liu2021pick}. In~\cite{zhang2022efraudcom}, Zhang et al. develop a competitive graph neural networks (CGNN)-based fraud detection system, that uses some notion of normal behavior as weak supervision information for the model to build a profile of fraudulent users. In 2018, Weber et al.~\cite{Weber2018} presented AMLSim\footnote{https://github.com/IBM/AMLSim}, a project intended to provide a multi-agent based simulator that generates synthetic banking transaction data together with a set of known money laundering patterns. The authors propose preliminary results showing that graph learning for AML is possible even in large sparse networks (18 nodes and 98 edges).

Zhong et al.~\cite{zhong2020financial} develop a framework that exploit RNNs to identify defaulter users. Here, financial transactions are modeled through attributed heterogeneous information networks (AHINs), thus leveraging multiple levels of user information to identify those that are exposed to a higher risk of default.
In general, user information - whenever available - is frequently used to enhance the performances of anomaly detection methods. This is true also for non-graph based methods. For instance, Jindal and Liu~\cite{jindal2008opinion} select a number of user features in order to identify anomalies in user-product review platforms; Lim et al.~\cite{lim2010detecting} perform a similar task to detect target-based spamming. Users’ behavioral patterns are explored in~\cite{hooi2016birdnest} and~\cite{wang2019fdgars}.

Given the lack of hand-labelled data to be used as a training set, many machine learning models are based on unsupervised anomaly detection~\cite{Chen2018}; techniques to generate realistic synthetic datasets have also been developed~\cite{altman2023realistic}. Other new approaches are based on zero-shot learning (and variations such as one-shot learning or few-shot learning) or meta-learning. For the task of money laundering detection, Pan~\cite{Pan2022} proposes a deep set algorithm based on both meta-learning and zero-shot learning. In the first meta-learning phase the model learns contrastive comparison to evaluate the membership of a query point against a given set of positive and negative samples. Then the model is further trained according to a sort of zero-shot learning with the aim of improving the accuracy of the model.

\subsection{Our contribution to the existing literature}
Many of these approaches show promise and effectiveness; however, this often comes at the expense of computational efficiency and interpretability of the results, falling short in meeting at least one of the requirements identified in Sec.~\ref{subsec:scope}. Our work distinguishes itself from the vast majority of the literature mentioned above by aiming to satisfy these requirements through the exploitation of the fundamental principles of graphs.

A notable advantage of NA is that, while many centrality algorithms can efficiently compute results for large graphs, their outcomes remain easily interpretable based on the chosen centrality measure. Dumitrescu et al.~\cite{dumitrescu2022anomaly} leverage basic node metrics and ego-networks as features for an anomaly detection algorithm, albeit on graphs smaller in size compared to those used in our work, and notably, they are labeled. Nevertheless, their results are highly encouraging for the application of simple NA-based methods in anomaly detection for financial transactions.

Another aspect that differentiates our approach from the majority of previous systems is its applicability to temporal networks at different resolutions. Many anomaly detection systems based on NA typically seek irregularities at a structural (static) level, overlooking changes observable in the network's evolution. Furthermore, nodes can be aggregated into groups (in our financial networks, transactions between countries, banks, or accounts can be aggregated). Observing behavioral differences at various resolutions allows domain experts to zoom in and out of their data when irregularities are observed, providing an additional layer of interpretability to their analysis.

\section{Notation, Measures, and Methods}\label{sec:methods}

\subsection{Basic notation}
\label{subsec:notation}

Networks are intuitive data structures that are suitable to model interactions between two or more agents. A financial transaction is the perfect example of such an interaction: two entities (e.g.,  individuals, bank accounts) exchange certain amounts of money in one or more transactions over time. If we assume that a timestamp is given for each transaction, then a \textit{directed weighted temporal network} provides a natural representation for this kind of data. 

If nodes are our entities, and links are the interactions (e.g., financial transactions) between them, the overall graph is simply $G=(N,L)$, where $N=\{\mbox{id}_1, \mbox{id}_2, \ldots, \mbox{id}_n\}$ is the set of nodes (or entities), each identified by an unique identifier, and $L = \{(i,j, w, t) : i, j \in N, w \in \mathbb{R}, t \in \mathbb{T}\}$ is the set of links (or edges), such that each quadruple denotes that at \textit{time} $t$ an \textit{interaction} from $i$ to $j$ with \textit{weight} $w$ has been recorded. Note that with $\mathbb{T}$ we denote the set of the timestamps, where a timestamp is just a date and a time. For instance, in the financial domain, the weight can be the overall \textit{amount} of money transferred from  $i$ to $j$ at time $t$. Also, timestamps and weights can be denoted respectively as $t(e)$ (or $t_{ij}$) and $w(e)$ (or $w_{ij}$) in terms of a given directed edge $e = (i, j)$, and they can be just two of a more general set of features, whose number is $f_L$.

Timestamps are defined in a given time period $\dot{T} = [t_0, t_\omega] \subset \mathbb{T}$, where $t_0$ and $t_\omega$ are the times of the first and the last recorded interactions in the dataset. Different time intervals $T_1, T_2, T_3, \ldots$ can be set within $\dot{T}$, such that a generic $T$ is defined by an interval $[t_x, t_y]$, where $t_x, t_y \in \dot{T}$ and $t_0 \le t_x < t_y \le t_\omega$. For example, if we assign the first and the last recorded timestamps of a given month respectively to the left and right extreme of the interval, we have a one month long time interval, that can simply be denoted by means of the month and the year, e.g.,  $T = \mbox{Dec}2023 = [\mbox{12/01/2023 @ 00:03:36am},\mbox{12/01/2024 @ 11:59:48am}]$, using UTC date time formatting. Let's also observe that $\dot{T}$ itself is also the largest possible time interval we can define. 
A \textit{temporal layer} $G_T$ of our overall graph $G$ is a snapshot of the system in a given time interval $T$: $G=(N,L_T)$, s.t. $L_T = \{e \in L : t(e) \in T\}$. 

Finally, each entity $i$ has $f_N$ number of features - such as demographics for an individual, or other strategic information - that can be used to create aggregations $G_T^\mathcal{F}$ of the graph $G$, or its temporal layers $G_T$, at different resolutions. For example, in the financial domain, we can set $T = \mbox{Dec}2023$ and $\mathcal{F}=\mbox{Country}$ is a graph layer representing all the transactions recorded in December 2023 between accounts aggregated by their countries, and an illustrative edge $e \in G_{\mbox{Dec}2023}^\mathcal{\mbox{Country}}$ can be $e = (\mbox{Italy},\mbox{France})$, where $w(e)$ is the overall amount of money sent from accounts in Italy to other accounts in France by means of different transactions occurred in that period.

\subsection{Centrality measures}\label{subsec:centralities}

In every graph, \emph{node centralities} can be measured in order to interpret the role of every node with respect to the others. This is done according to some specific definition of centrality, that can provide different insights on the role of every entity belonging to the network. For example, we can evaluate the importance of every node in the most intuitive way by simply checking its \emph{degree centrality}, i.e., by counting the number of neighbours. A node with many neighbours will most likely be very ``popular'' in our system, as opposed to a node with very few connections, and will be in the top positions of a ranking based on the values of the degree centrality. This simplistic hypothesis can be refined as required, in order to evaluate a node's importance by taking into account other features of the connections other than the pure number, such as quality (that can be defined differently in different contexts), frequency over time, and many others. The most common node's attributes are the following:
\begin{definition}[Predecessors, successors, and neighbors]
Given node $j$, all the nodes $i$s for which exist a direct link $(i,j) \in L$ are called $j$'s \textit{predecessors}, and their set is denoted as $N^-_j$. Similarly, given node $i$, all the nodes $j$s for which exist a direct link $(i,j) \in L$ are called $i$'s \textit{successors}, and their set is denoted as $N^+_i$. The union of the predecessors and successors of $i$ is the set of \textit{neighbors} of node $i$, and it is denoted by $N_i$.
\end{definition}
\begin{definition}[Indegree, Outdegree, and Degree]
The numbers of predecessors, successors and neighbors of $i$ are respectively the \textit{indegree} $k^-(i)$, \textit{outdegree} $k^+(i)$, and \textit{degree} $k(i)$.
\end{definition}

If the graph is also weighted, we can define the following measures:
\begin{definition}[Instrength, Outstrength, and Strength]
Given a node $i$, the instrength $s^-(i) = \sum_{j : N^-_j } w_{ji}$ is the sum of the incoming links' weights, the outstrength $s^+(i) = \sum_{j : N^+_i } w_{ij} $ is the sum of the outgoing links' weights, and the strength is the sum of all the undirected links' weights $s(i) = \sum_{j : N_i } w_{ij}$.
\end{definition} 

Some nodes are particularly interesting in terms of connectedness:
\begin{definition}[Sources, Sinks, and Singletons]
A node $i$ with $k^-(i) = 0$ is called a source, as it is the origin of each of its outgoing links. Similarly, a node $i$ with $k^+(i) = 0$ is called a sink, since it is the end of each of its incoming arcs. If a node $i$ is both a sink and a source, i.e., $k^-(i) = k^+(i) = 0$, then it is disconnected from the rest of the network and it is called a singleton.
\end{definition}

In our work, we focus on a set of node centrality measures that evaluate different aspects of the importance of each node, considering not only the pure number of connections, but also their \textit{strength}, as well as their \textit{quality}. The centrality measures considered, that are synthesised in Tab.~\ref{tab:centralities}, are defined as follows:

\begin{definition}[Indegree, Outdegree, and Degree centralities]
Similarly, indegree (outdegree) centrality $d^-(i) = k^-(i)/n$ ($d^+(i) = k^+(i)/n$) of node $i$ is the normalised version of node's indegree (outdegree). The degree centrality $d(i)$ of node $i$ is the normalised version of node's degree: $d(i) = k(i)/n$. 
\end{definition}

Indegree centrality, as well as the degree centrality in undirected networks, is a very simple measure of popularity of a node.
Conversely, outdegree centrality is also called the branching factor because it is a measure of the extent to which a node is connected to other nodes in the network, or the number of other entities to whom the node can spread information.

Quite naturally, (in/out)degree centralities can be extended to (in/out)strength centralities: in these cases, we just sum the (incoming/outgoing) links' weights, normalizing by the sum of all the links' weights if we need a measure in the $[0,1]$ interval.

A node that is closer to all the other nodes on average has its own centrality in a network: in fact, closeness centrality of a node is defined as the average distance from it to all the other nodes in the network. In order to get a value that is in the range of $(0,1]$, we can compute the reciprocal of the average:  
\begin{definition}[Closeness and Harmonic Closeness]
The (normalized) closeness $c(i)$ and the harmonic closeness $\bar{c}(i)$ of a node $i$ are respectively calculated as \[c(i) = \frac{|N|-1}{\sum_{j\neq i}l_{ij}}, \bar{c}(i) = \sum_{j \in N}\frac{1}{l_{ij}},\] where $l_{ij}$ is the shortest path length (distance) between node $i$ and node $j$.
\end{definition}
Note that Harmonic closeness is just a workaround to calculate the measure in disconnected graphs, where distances between nodes in different components are infinite. 

Instead of popularity or topological centrality of a node, we may be interested to \emph{bridges}, that are nodes crossed by shortest paths more often than other nodes in the networks. 

\begin{definition}[Node Betweenness]
The betweenness $b(i)$ of node $i$ is \[b(i) = \sum_{s \ne t \ne i} \frac{\rho_{st}(i)}{\rho_{st}},\] where $\rho_{st}$ is the number of shortest paths from $s$ to $t$, and $\rho_{st}(i)$ is the number of shortest paths from $s$ to $t$ passing through  $i$.
\end{definition}

Another set of measures is made of variants of the so called \emph{eigenvector centrality}, exploiting the spectral properties of a given graph. \emph{Katz centrality}, \emph{PageRank} and \emph{Hubs and Authorities scores} (calculated by the \emph{HITS} algorithm) are examples of such measures. This family of centralities aims to measure the influence of a node in a network. These measures are natural extensions of simple degree centrality, whose major limitation is to assign to all node's neighbors the same importance. With eigenvector centrality, relative scores are assigned to all nodes in the network based on the concept that connections to high-scoring nodes contribute more to the score of the node in question than equal connections to low-scoring nodes. A high score means that a node is connected to many nodes who themselves have high scores. Here, we recall the PageRank, originally introduced in~\cite{brin1998anatomy},~\cite{page1999PageRank} for ranking web page results by their network's importance and the Hub and Authority scores~\cite{kleinberg1999hubs}, calculated by the so called \textit{hyperlink-induced topic search} algorithm (also known as HITS), since both methods works well in directed graphs.

\begin{definition}[PageRank]
The PageRank $\mbox{pr}(i)$ of node $i$ is calculated by:
     \[\mbox{pr}(i) = \alpha\sum_{j \in N^-_j}\frac{\mbox{pr}(j)}{k^+(j)} + \frac{1-\alpha}{n},\]
     where $\alpha$ is a positive constant in the range $(0,1)$ called the damping factor.
\end{definition}

PageRank is higher for nodes that have a higher probability to be reached, sooner or later, by means of a \emph{random walk} process, regardless of the node selected as a starting point. The damping factor $\alpha$ represents the probability, at any step of the random walk, that the walker will continue following links; instead, $(1-\alpha)$ is the probability that they jump to any random page. It should also be noted that the PageRank algorithm can be extended to weighted graphs, even if we do not report here its reformulation for the sake of simplicity.

In some networks it is appropriate also to assign a node high relevance's value if it points to others with high relevance, and this could be particularly useful for directed networks. We can define two different types of measures, i.e., the \emph{authority} score and the \emph{hub} score, which quantify nodes' prominence in the two roles, as a ``receiver'' or as a ``provider'' of information. 
The HITS algorithm introduced and implemented such idea, and it has been originally applied to Web retrieval tasks, as well as measuring inbound and outbound behaviors in other domains (e.g., to study the world trade network~\cite{Deguchi2014}, or the brain drain phenomenon in the scientific migration network~\cite{urbinati2021measuring}).

The HITS algorithm assigns each node $i$ an authority score $a_i$ and hub score $h_i$. The defining characteristic of a node with high authority score is that it is pointed to by many hubs, and the defining characteristic of a node with high hub score is that points to many authorities. Hubs and authorities can be asymmetrical, just pointing out to different nodes' characteristics.
In light of these considerations, we can define the

\begin{definition}[Authority and Hub Scores]
    The authority score $a_i$ of a node $i$ is defined to be proportional to the sum of the hub centralities of its predecessors; similarly, the hub score $h_i$ of a node $i$ is proportional to the sum of the authority centralities of the nodes it points to. 
    \[a(i) = \alpha\sum_{j \in N^-_i}w_{ji}h(j), h(i) = \alpha\sum_{j \in N^+(i)}w_{ij}a(j),\]
    where $\alpha$ is a constant $\in (0,1)$. 
\end{definition}

The PageRank and the HITS algorithms are iterative, and they stop when the values stabilize. This stability is guaranteed, and reached by means of few steps by the so called power method. Other information can be found in the original papers.

\begin{table}[ht]
\centering
\resizebox{\textwidth}{!}{%
\begin{tabular}{|l|l|l|} 
\hline
\textbf{Measure} & \textbf{Intuitive interpretation} & \textbf{Description} \\ \hline
Indegree $d^-(i)$ & Node $i$'s popularity & The number of $i$'s incoming connections.\\ 
Outdegree $d^+(i)$ & Node $i$'s branching factor & The number of $i$'s outgoing connections.\\ 
Instrength $s^-(i)$ & Node $i$'s weighted popularity & The total weight of $i$'s incoming connections.\\ 
Outstrength $s^+(i)$ & Node $i$'s weighted branching factor & The total weight of $i$'s outgoing connections.\\ 
Closeness $c(i)$ & Node $i$'s network topological centrality & The smaller the distances to other nodes, the greater its value.\\
Betweenness $b(i)$ & Node $i$'s importance as a bridge & The higher the number of shortest paths passing through $i$, the greater its value.\\
PageRank  $\mbox{pr}(i)$ & Node $i$'s network relevance & The probability that a random walker will eventually reach $i$.\\ 
Hub Score $\mbox{h}(i)$ & Node $i$'s network relevance as a sender & The importance of $i$ as a sender, based on the importance of its successors.\\ 
Authority Score $\mbox{a}(i)$ & Node $i$'s network relevance as a receiver & The importance of $i$ as a receiver, based on the importance of its predecessors.\\ 
\hline
\end{tabular}
}
\caption{A list of centrality measures commonly used to generate node rankings.}\label{tab:centralities}
\end{table}

\subsection{Node rankings}
\label{subsec:noderankings}
\begin{figure}[h!]
\centering
\includegraphics[width=0.7\textwidth]{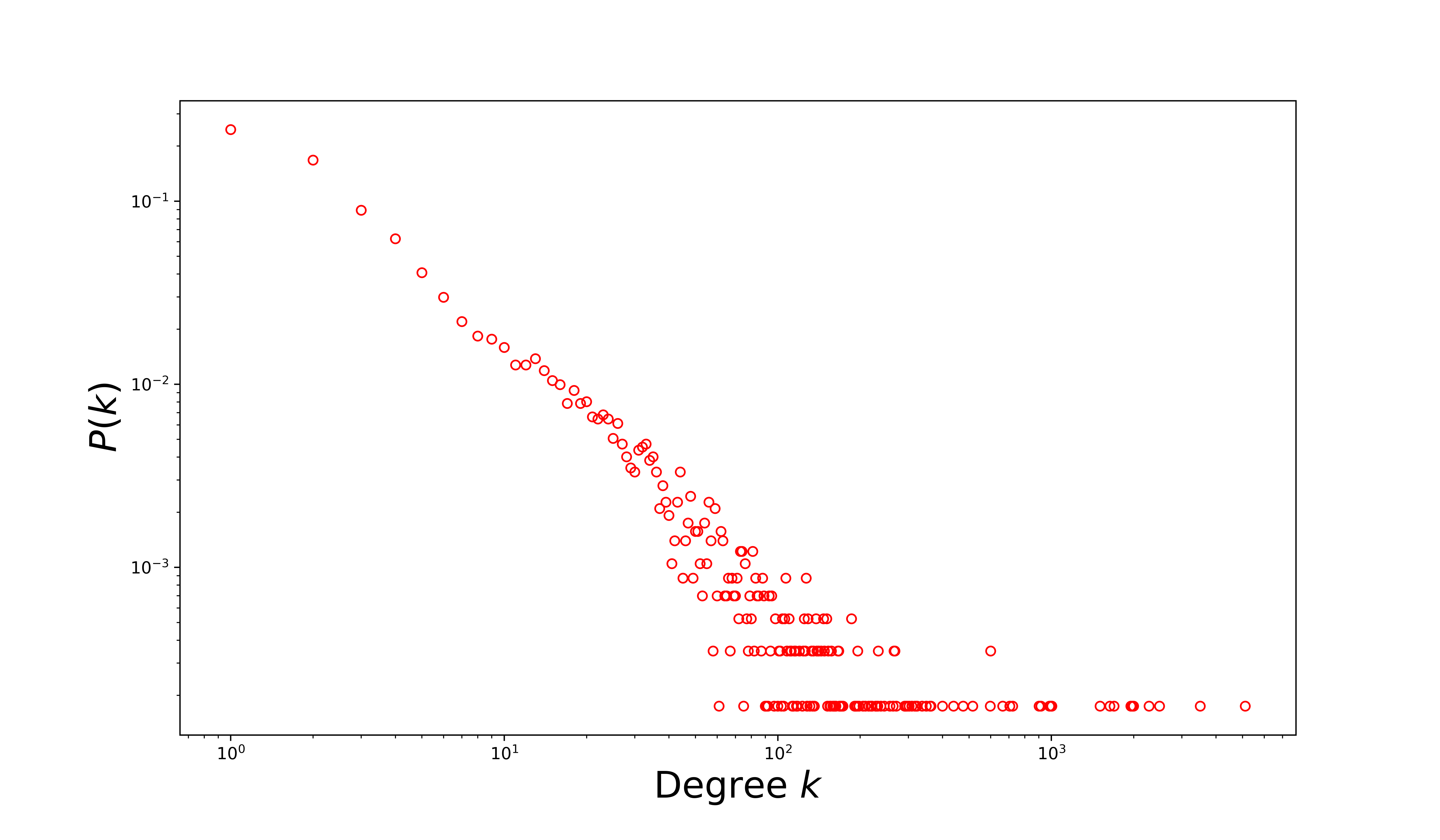} 
\caption{Degree distribution for $G^\text{BIC}_{\text{Feb2022}}$.}
\label{fig:degree_distr}
\end{figure}

All the above measures can be used to rank nodes accordingly: we use ranks' intervals to identify classes of nodes (or links) with comparable properties, to assume that, under given circumstances, they have similar roles in the network and in its dynamics. Such statistical approach is commonly applied by plotting centrality distributions, that very often show heavy-tailed shapes, corresponding to highly heterogeneous classes of nodes. Focusing on degree centrality, the heterogeneity of the distribution leads to the emergence of nodes with very large (in)degree~\footnote{Please, observe that in the literature this is often referred as the ``emergence of hubs phenomenon''~\cite{barabasi1999}, but the term ``hub'' is used there to indicate a high (in)degree node, whereas a node with high hub score, according the HITS algorithm, has a different meaning.}, making statistically meaningless the adoption of the average $\langle k \rangle$ as a scale of degree centralities. In other words, the average degree should not be used as a good estimate of a randomly selected node's degree in many real large scale networks. An example of an heterogeneous degree distribution can be see in Fig.~\ref{fig:degree_distr}, where we plot the degree distribution of one of the graphs built upon the dataset that we will introduce in Sec.~\ref{subsec:dataset}.

In our approach, we will compute node rankings based on various measures to track their evolution over time. Consequently, we require methods to compare different rankings. 

For this purpose, we will employ the classic Spearman's rank correlation coefficient~\cite{spearman1904}, denoted as Spearman's $\rho$. This non-parametric measure assesses the rank correlation between two sets of data.

\begin{definition}
The Spearman correlation coefficient is defined as the Pearson correlation coefficient between two rank variables $r_x$ and $r_y$:
\[\rho(r_x,r_y) = \frac{\mbox{cov}(r_x,r_y)}{\sigma(r_x)\sigma(r_y)} \]
where $\mbox{cov}(r_x,r_y)$ is the covariance of the rank variables, and $\sigma(r_x),\sigma(r_y)$ are the standard deviations of the rank variables.
\end{definition}

Another classic measure is the Kendall rank correlation coefficient~\cite{kendall1938new}, commonly denoted as Kendall's $\tau$ coefficient.  As Spearman's $\rho$, Kendall's $\tau$ is a measure of rank correlation: the similarity of the orderings of the data when ranked by each of the quantities. More precisely, it counts the number of concordant and discordant pairs, and the difference between the two quantities by the total number of possible pairs. 

Both correlation coefficients have values in the range $[-1, 1]$, ranging from full disagreement to full agreement between the two series, with value equal or very close to zero when they are independent.

\subsection{Evaluation of an outlier detection system in terms of relevance}\label{subsec:rankeval}
As mentioned in Sec.~\ref{subsec:scope}, we approach the anomaly detection task as a problem of ranked information retrieval. In this context, classic evaluation measures such as Precision, Recall, or F-measure, calculated on unordered sets of items, are not suitable. Due to our assumptions and requirements, even the use of a precision/recall graph is precluded, as we lack a reliable ground truth, rendering Recall impossible to calculate.

Nevertheless, given requirement \textbf{R2}, our system should provide experts with a concise list of potentially relevant outliers, so that we can still assess our system estimating the values of well established metrics like \textit{Precision-at-K} (\textit{P@K}), \textit{Average Precision} (\textit{AP}), and other relevant measures. P@K evaluates the precision of identified anomalies within the top-K ranked instances, providing insights into the relevance and accuracy of these detections. It measures the proportion of correctly ranked anomalies within a specified threshold, facilitating an understanding of model efficacy. The metric is defined as follows:

\begin{equation}
    \text{P@K}(r)= \dfrac{\text{tp@K}(r)}{K}
\end{equation}

where $K$ is a given position of the ranking, and \textit{tp@K} is the number of outliers marked as relevant (i.e., true positives) in top-$K$ by the domain expert.  Therefore, a $\text{P@5} = 0.6$ would mean that, in the first 5 positions of the rankings, 3 entries out of 5 are relevant.

Moreover, the \textit{average P@K} is defined as the average of the \textit{P@K} considering all the positions in the ranking where we have a relevant outlier.  Complementing P@K, Average Precision considers the precision across various retrieval cutoffs, providing a nuanced assessment of the model's performance throughout the entire ranking list. 

Finally, since we do not always have information on all the true negatives, in some experiments we are not able to compute the recall; nonetheless, we can define a  recall-like measure $R^*$ 

\begin{equation}
    R^*(r) = \dfrac{\text{tp}(r)}{\text{tp}^*}
\end{equation}

where $\mbox{tp}^*$ is the total number of outliers marked as relevant by the expert among the set returned by the algorithm.

\section{Our Proposal}
\label{sec:proposal}

\subsection{Problem conceptualization}\label{subsec:probdef}

As noted by Savage et al. in~\cite{SAVAGE201462}, when detecting anomalies in social networks, our focus is on identifying unexpected patterns of interactions between individuals within the network. Consequently, network anomalies can be succinctly defined as ``patterns of interaction that significantly differ from the norm,'' aligning with the definitions provided by Chandola et al.~\cite{chandolaetal2009}, and Hodge and Austin~\cite{hodgeaustin2004}. Broadening the perspective, rather than limiting our consideration to patterns in their typical network analysis connotation — such as motifs and paths — we can explore and identify more general forms of regularities.

Hence, we present a comprehensive framework for distinguishing regularities from anomalies. In this context, we posit that our graph exhibits a certain regular property, and any observation deviating from this property is deemed an outlier and, potentially, as a system anomaly. Our objective is to, with a well-defined characterization of this property and empirical validation that it holds for our graph, identify outliers that do not conform to this specified property.

We can check if the given property holds within an individual graph and/or if it is consistently present over time: in the latter case, we will have a temporal property $\overline{\mathcal{P}}$. For example, we observe that the graph's degree distribution follows a power law at the interval $T_0$; we may wonder if the same property holds at $T_1, T_2, \ldots$. If this property holds for $G_{T_0}, G_{T_1}, \ldots, G_{T_{x-1}}$, but it does not hold for $G_{T_x}$, we found a graph-level anomaly, that could be investigated further. 

In our scenario, intuitively, we are not focused on properties that are strictly invariant (i.e., the probability of observing an anomaly is extremely unlikely). Instead, we are searching for properties where outliers ``sometimes'' deviate from the general behavior. Simultaneously, we are not interested in extremely heterogeneous properties, as defining a ``normal'' behavior in such cases could be challenging. It's worth noting that some seemingly natural attributes may not necessarily serve as good candidates for defining the properties we seek. For instance, node strength and degree distributions are known to be highly heterogeneous in many real-world networks, and the probability of a node having strength and degree much higher than the mean is significantly higher than with normal distributions.

\begin{definition}[Structural Property]
     Given a graph $G=(N,L)$, and a node $i$ of $N$, a structural property $P(G, i)$ is a true/false condition that is defined on features and measures of $i$. We say that $P$ holds for $i$ if the condition is true. 
\end{definition}

This definition is intentionally broad, adaptable to various graph metrics. For instance, our interest might lie in computing the ranking of nodes based on indegree and PageRank. This entails having two rank variables, $r(d^-,i)$ and $r({\mbox{pr},i)}$, denoting the position of node $i$ in the two rankings, respectively. 

It is worth noting that we can also define structural properties that are not based on strict equalities. For example, we can relax the structural property defined above to check if a node occupies \textit{approximately} the same position in both rankings: $P(G, i) = (r(d^-,i) \approx r({\mbox{pr}},i))$. We can easily describe mathematically this property in terms of some approximation threshold, so that if $N = \{1, 2, \ldots, 5\}$, and $r(d^-,2) = 3$, and $r({\mbox{pr}},2) = 2$, then we can say that $r(d^-,2) \approx r({\mbox{pr}},2)$, and that $P$ holds for node $i=2$. Conversely, if $r(d^-,3) = 1$ and $r({\mbox{pr}},2) = 5$, then $r(d^-,3) \not\approx r({\mbox{pr}},3)$, and $P$ does not hold for node $i=3$.

Once we defined one or more structural properties, we want to check if they either hold or do not hold for a substantial majority of nodes in a graph. 

\begin{definition}[Graph Structural Property]
Given a graph $G=(N,L)$, and a structural property $P$, a graph structural property $\mathcal{P}(G, P)$ is valid if $P(G,i)$ significantly holds for nodes $i \in N$.
\end{definition}

For example, we may wish to verify if every node occupies approximately the same position in both rankings. By defining our structural property as $P(G, i) = (r(d^-,i) \approx r({\mbox{pr}},i))$, we can check its validity for every node in the graph by solving the following logical equation: $\forall_{i \in N} P(G, i)$. If the latter is true, it means that the two rankings are approximately the same. However, the broadness of the above definition allows us to do something more: we can also check \textit{how much} the rankings are similar or dissimilar using rank correlation coefficients (like Spearman's $\rho$ and Kendall's $\tau$ we recalled in Sec.~\ref{subsec:noderankings}). 
For example, with $N = {1, \ldots, 5}$ and node rankings calculated according to indegree and PageRank, represented as $r(d^-) = [3, 2, 4, 1, 5]$ and $r({\mbox{pr}}) = [2, 3, 4, 1, 5]$, we can assert that the graph structural property built upon $P(G, i) = (r(d^-,i) \approx r({\mbox{pr}},i))$ is valid with $\rho = 0.9$ and $\tau = 0.8$. Conversely if $r(d^-) = [3, 2, 4, 1, 5]$ and $r({\mbox{pr}}) = [2, 3, 4, 5, 1]$ is not valid, because $\rho = -0.6$ and $\tau = 0.4$: even if only two indices switched they rank positions with each other as before, their fluctuations were much higher (from 5 to 1 and from 1 to 5), and rank correlation coefficients are sensible to those changes.

In a time-varying network, we may be interested to properties that could be checked \emph{across} different temporal layers. Hence, when we observe how a metric changes over time, we need to define a temporal property:

\begin{definition}[Temporal Property]
Given a graph $G = (N,L)$, two distinct time intervals $T_x$ and $T_y$, and a measure or feature $f$ associated with a generic node $i$, a temporal property $\overline{P}(G, T_x, T_y, f, i)$ is a condition that evaluates to true if $f(i)$ remains approximately unchanged when computed for $G_{T_x}$ and $G_{T_y}$. We say that $\overline{P}$ holds for $i$ if the condition is true.
\end{definition}

If the feature we are considering is $f(i) = r(d^-,i)$, that is the position of $i$ in a ranking based on the indegree, then we can build on $f$ (the indegree) a temporal property. If the position of $i$ in the ranking remains approximately unchanged between $T_x$ and $T_y$, i.e., $r_x(d^-,i) \approx r_y(d^-,i)$, then  $\overline{P}(G, T_x, T_y, f, i)$ holds for node $i$.

Finally, analogously as we did with the graph structural property, we need to formalize the idea that a temporal property should be tested on every node of the graph, and considered valid if it significantly holds for them:
\begin{definition}[Graph Temporal Property]
Given a graph $G=(N,L)$, two distinct time intervals $T_x$ and $T_y$, a measure or feature $f$ associated with a generic node $i$, and a temporal property $\overline{P}$, a graph temporal property $\overline{\mathcal{P}}(G, \overline{P})$ is valid if $\overline{P}(G, T_x, T_y, f, i)$ significantly holds for nodes $i \in N$.
\end{definition}

For example, with $N = \{1, \ldots, 5\}$, indegree based node rankings in two different $T_x$ and $T_y$ are $r_x(d^-)$ and $r_y(d^-)$, we build graph temporal property $\overline{\mathcal{P}}(G, \overline{P})$ in terms of temporal property $\overline{P}(G, T_x, T_y,d^-, i) = (r_x(d^-,i) \approx r_y(d^-,i))$. We can decide to consider $\overline{\mathcal{P}}$ valid if Spearman's or Kendall's tau correlation coefficients are greater than $0.5$ (stronger or weaker threshold can be set depending on the domain).

The conceptualization introduced in this section assists us in directing our analysis towards temporal properties and node metrics that exhibit stability over time. Stability, in this context, implies that the majority of nodes do not (significantly) alter their values. Hence, a valid graph temporal property can be defined in terms of stable node's metrics. This approach enables us to promptly identify the minority of nodes that deviate from the norm. 

An additional observation is necessary here: the conceptualization pertains to measures and features at the node level; we can readily extend our definitions to also encompass links. This extension is particularly relevant in certain domains where the focus may be on the interactions themselves, their overall stability, and the minority of links that deviates from the norms. Although the formal conceptualization can be also extended to other kind of analyses that are not necessarily focused on single nodes, the rest of the paper will not mention this aspect for the sake of clarity.


\subsection{A dance of nodes: the anomaly detection pipeline}\label{subsec:pipeline} \label{subsec:methods_pak}
In this section we describe a general system that returns a top-K list of nodes ranked by their measured deviation from a norm.  Specifically, our methodology employs a
suite of network centrality measures to be calculated on every node. By monitoring the evolution of centrality-based
node rankings over time, our approach effectively identifies nodes whose changes are significantly relevant. 

The pipeline includes the following steps:
\begin{enumerate}
    \item \textbf{Building the graphs}: let's start from representing the data with a graph $G=(N,L)$, whose entities are nodes, and links are recorded interactions between nodes. Each  link $(i,j)$ has a timestamp $t_{ij} \in \dot{T}$ and one weight $w_{ij}$. Then, we group nodes in $N$ by feature $\mathcal{F}$, and finally we define the duration of a time interval (e.g., a month) to extract all the temporal layers $G_{T_0}^\mathcal{F}, G_{T_1}^\mathcal{F}, \ldots$. For the sake of simplicity, we omit the aggregation feature $\mathcal{F}$ from the following notations.
    \item \textbf{Computing centrality metrics}: on each graph, we compute (a selection of) the centrality metrics reported in Table~\ref{tab:centralities}. Hence, for every $i \in N_{T_x}$, we calculate its generic metric $m_x(i)$, where $m \in \{d^-, d^+, s^-, s^+, c, b, \mbox{pr}, h, a\}$, and $x$ denotes the time interval $T_x$, e.g., indegree $d_x^-(i)$, outdegree $d_x^+(i)$, instrength $s_x^-(i)$, outstrength $s_x^+(i)$, and so on.
    \item \textbf{Building the centrality-based node rankings}: for each graph $G_{T_x}$, we produce node rankings according to every network centralities we computed at the previous step. If a ranking based on metric $m$ can be generally denoted as $r(m)$, where $m \in \{d^-, d^+, s^-, s^+, c, b, \mbox{pr}, h, a\}$, we can also denote with $r_x(m)$ the ranking based on metric $m$ at time interval $T_x$; $r_x(m, i)$ denotes node $i$'s position according to $r_x(m)$, e.g., $r_x(d^-, i) = 1$ means that $i$ is the top ranked node by indegree in $G_{T_x}$.
    \item \textbf{Defining the temporal properties}: for each time intervals pair $T_x$ and $T_y$, we define temporal properties $\overline{P}(G, T_x, T_y, r(m), i)$ in terms of the centrality-based node rankings calculated in the previous steps, i.e., 
    $\overline{P}(G, T_x, T_y, r(d^-), i)$ holds if $r_x(d^-, i) \approx r_y(d^-, i)$. 
    \item \textbf{Stability check}: as argued in Sec.~\ref{subsec:probdef}, we need to verify if the graph temporal properties $\overline{\mathcal{P}}$ are valid, e.g., if the temporal property is $\overline{P}(G, T_x, T_y, r(d^-), i)$ as defined above, we need to check if $r_x(d^-, i) \approx r_y(d^-, i)$ holds significantly for nodes $i \in N$. This can be done by calculating the rank correlation coefficients of $r_x(m)$ and $r_y(m)$, given time intervals $T_x$ and $T_y$. 
\end{enumerate}

If the stability check is passed, we can exploit the validity of the graph temporal property to adopt the temporal fluctuations of nodes' positions in centrality based rankings as a criterion for sorting nodes.

\begin{enumerate}
\setcounter{enumi}{5}
    \item \textbf{Time-varying comparison}: we can compare node rankings at different time intervals $T_x$ and $T_y$. For each node $i$, we calculate the \textit{residual} $\delta$, namely the difference in its positions in two centrality-based rankings, as follows: $\delta(T_x, T_y, r(m), i) = r_x(m,i) - r_y(m,i)$, where $r_x(m,i)$ and $r_y(m,i)$ are node $i$'s ranks according to metric $m \in \{d^-, d^+, s^-, s^+, \dots\}$, calculated respectively for $G_{T_x}$ and $G_{T_y}$. Note that if the residual is zero, it indicates that the node's rank did not change from $T_x$ to $T_y$; if it is negative, it means that the node has a lower rank in $T_y$ than in $T_x$; if it is positive, it means that the node has a higher rank in $T_y$ than in $T_x$. Consequently, we sort our nodes by residual for every metric $m$ used to create our rankings. It is important to note that we can sort by $|\delta|$ if we are indifferent to whether nodes gained or lost positions in their rankings, and we want just to sort by the magnitude of such change. Alternatively, we can sort in ascending or descending order based on whether we want to prioritize nodes that respectively lost or gained positions in their ranks from $T_x$ to $T_y$.
    
    \item \textbf{Node filtering} (optional): an optional step of node filtering can be taken in order to exclude from the calculation of the final output some nodes matching a simpler rule of ``irrelevance'' given by the domain expert. Additionally or in alternative, an unsupervised machine learning based outlier detection algorithm - such as OneClassSVM~\cite{oneclasssvm1999} or Isolation Forest~\cite{liu2008isolation} - can be applied in order to filter out irrelevant nodes from the final output;
    \item \textbf{Returning the top-K anomalies}: at this stage we have collected a number of ordered lists, based on as many network centrality measures, that sort nodes according to their \textit{residual}, namely their deviation from the stability's condition. Depending on the specific use case and on the needs of the user, we can choose to adopt, as a final output of anomalies, a list generated by one of the centrality measures in step 6/7. Anyway, depending on the context and on the chosen centralities, it can happen that these outputs are heavily different, meaning that different nodes are flagged as outliers according to different definitions of importance. If one needs to incorporate the knowledge coming from more than one centrality metric in a single list, then some \textit{mixed strategy} can be implemented to obtain the desired output. For instance, top-K anomalies can be selected 
    from all the top-K lists obtained so far returning a ``meta'' top-K list that is the result of the application of some heuristic. It is worth recalling that such K should be kept as smaller as possible, after requirement \textbf{R2} (see Sec.~\ref{subsec:scope}).
\end{enumerate}

In some domains, we expect that the final ranking strategies based on only one network metric based ranking will perform much better than the others in terms of precision@k: in these cases, we can decide to embed only that metric in our pipeline, neglecting the others. In some other domains, more than one network metric can contribute to detect fresh and relevant anomalies: we will see an example in Sec.~\ref{sec:casestudy}.



   

\subsection{A pipeline's illustrative example and node rankings evolution charts}\label{subsec:rec}
\begin{figure}
  \centering

  \begin{subfigure}{0.45\textwidth}
    \centering
    \includegraphics[width=\linewidth]{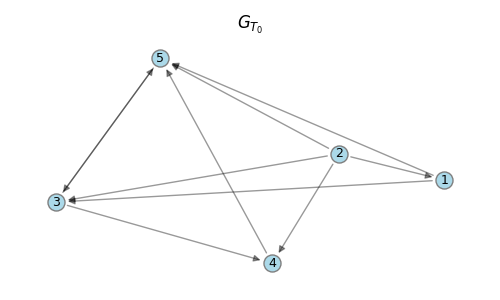}
    \caption{Graph's temporal layer $G_{T_0}$}
    \label{fig:gt0}
  \end{subfigure}%
  \hfill
  \begin{subfigure}{0.45\textwidth}
    \centering
    \includegraphics[width=\linewidth]{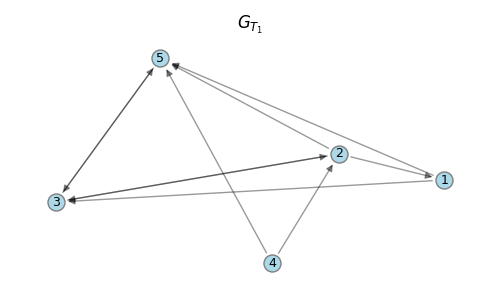}
    \caption{Graph's temporal layer $G_{T_1}$}
    \label{fig:gt1}
  \end{subfigure}

  \medskip

    \begin{subfigure}{0.45\textwidth}
    \centering
    \begin{tabular}{|c||c|c||c|c||c|}
      \hline
        &  \multicolumn{2}{c||}{$G_{T_0}$} & \multicolumn{2}{c||}{ $G_{T_1}$} & \\
        Node & indegree & rank & indegree & rank & $\delta$ \\
      $i$ &  $d_0^-$ &  $r_0$ &
       $d_1^-$ &  $r_1$ & $r_0 - r_1$\\
      \hline
      1 & .2 & 4 & .2 & 4 & 0 =\\
      2 & .0 & 5 & .4 & 3 & +2 $\triangle$\\
      3 & .6 & 2 & .6 & 2 & 0 =\\
      4 & .4 & 3 & .0 & 5 & -2 $\triangledown$\\
      5 & .8 & 1 & .8 & 1 & 0 =\\
      \hline
    \end{tabular}
    \caption{Evolution of nodes' indegree and rank}
    \label{fig:indegrees_and_ranks}
  \end{subfigure}%
  \hfill
  \begin{subfigure}{0.45\textwidth}
    \centering
    \includegraphics[width=\linewidth]{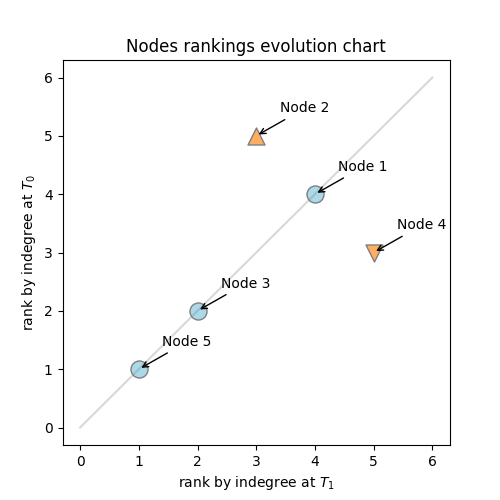}
    \caption{Node rankings evolution chart; anomalies colored in light orange, regular nodes in light blue.}
    \label{fig:evolution_chart}
  \end{subfigure}

  \caption{The temporal network based anomaly detection pipeline applied to a toy example.}
  \label{fig:examplepipeline}
\end{figure}

We provide here an illustrative example, introducing node ranking evolution charts to give a more intuitive and visual interpretation of the top-K anomalies selection process (see Fig.~\ref{fig:examplepipeline}).

Let's suppose that our data can be represented by a simple graph $G$ with 5 nodes and two temporal layers $G_{T_0}$ (Fig.~\ref{fig:gt0}) and $G_{T_1}$ (Fig.~\ref{fig:gt1}), accomplishing step (1) of our pipeline. We can calculate our centrality metrics for both  $G_{T_0}$ and $G_{T_1}$ at step (2), and then we can rank nodes accordingly at step (3). We focus here only on indegree $d^-$: the table in Fig.~\ref{fig:indegrees_and_ranks} show, for every node $i$ the indegree centralities $d_0^-(i)$ and $d_1^-(i)$, and the indegree based node ranking $r_0(d^-,i)$ and $r_1(d^-,i)$. The temporal property  $\overline{P}(G, T_0, T_1, r(d^-), i)$ is set to the condition $r_0(d^-,i) \approx r_1(d^-,i)$ at step (4). Setting a (very conservative) approximation threshold to $\pm 1$, the $\overline{P}$ holds for nodes $\{1, 3, 5\}$. After step (5) the graph temporal property $\overline{\mathcal{P}}$ defined in terms of $\overline{P}$ is validated, since our rank correlation coefficients are largely positive, e.g., Spearman's $\rho = 0.6$, and Kendall's $\tau = 0.4$. 

Once the stability check is passed, we can continue the final part of the process, in order to return the top-K anomalies. Looking at  Fig.~\ref{fig:indegrees_and_ranks}, at step (6) we execute the time-varying comparison between ranks at $T_0$ and $T_1$: it shows clearly that $|\delta| > 1$ for nodes $2$ and $3$. In particular, node $2$'s  $\delta$ is positive ($=+2$), that means that it \textit{gains} a higher rank from $T_0$ to $T_1$. On the contrary, node $4$ \textit{loses} two positions in the rank, so $\delta$ is negative. These two anomalies are returned to the domain expert. It should be noted that, in this example, step (7) is trivially skipped, and step (8) returns the same ordered list as in step (5).

An intuitive way to visualize and interpret what happens from step (6) to (8) is to use a \textit{node rankings evolution chart} (in the rest of the paper referenced as \textit{REC}) shown in Fig.~\ref{fig:evolution_chart}, a simple scatter-plot where values in the $x$ axis are the nodes' ranks at one time interval (e.g., $T_1$), and values in the $y$ axis are the nodes' ranks at the other time interval (e.g., $T_0$). If we draw the identity line $y = x$, we expect that the temporal property holds for nodes that will be plotted along (or very close by) the identity line. Conversely, outliers are progressively more and more distant from the identity line. 

\begin{figure}[ht!]
\centering
    \begin{subfigure}{0.45\textwidth}
        \centering
        \includegraphics[width=\linewidth]{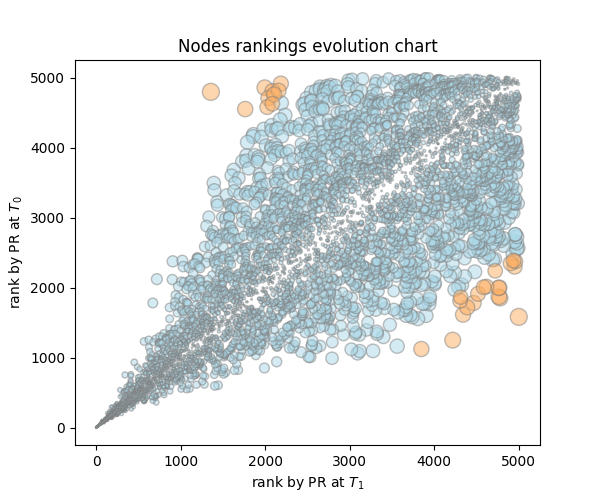}
        \caption{Anomalies are the top-30 of a node list sorted by the absolute value of the residual.}
        \label{fig:strategy1}
    \end{subfigure}%
    \hfill
    \begin{subfigure}{0.45\textwidth}
        \centering
        \includegraphics[width=\linewidth]{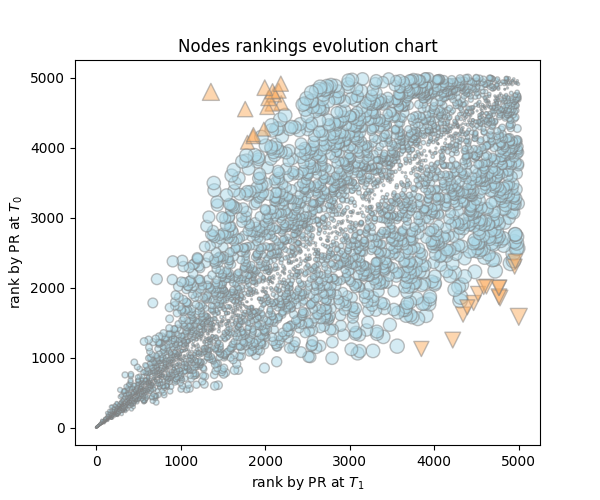}
        \caption{Anomalies are the top-15 `positive' and the top-15 `negative' outliers.}
        \label{fig:strategy2}
    \end{subfigure}
\caption{The evolution charts shown above has been plotted after that PageRanks have been measured for all the nodes in the graph, and node rankings in time intervals $T_0$ and $T_1$ are  calculated afterwards. Markers' sizes are drawn proportionally to the absolute value of the residual. Finally, in red, you find K=30 outliers}
\label{fig:examplepipeline2}
\end{figure}

If the graph is large, we usually have much more than K (with small K because of requirement \textbf{R2}) outliers, i.e., nodes for which the temporal property does not hold. In that case we can order the list of outliers by their residual w.r.t. the identity line. This is analogous to sort nodes by $|\delta|$, and then return the top-K elements of such a list. For example, Fig.~\ref{fig:examplepipeline2} shows the nodes ranking evolution charts obtained by calculating PageRank for every node of a randomly generated directed graph $G_{T_0}$, and of a variant $G_{T_1}$, obtained by reshuffling $10\%$ of randomly selected edges in $G_{T_0}$\footnote{The example's original graph is a Barabasi-Albert random graph with 5,000 nodes, and 5 new links attached at each step of the generative process.}. Then, nodes are ranked in both graphs according to their PageRanks. As before, we can calculate, for each node, its $\delta$ between the two ranks.

Going directly to step (8) of the pipeline, we aim to return $K=30$~\footnote{Other small values for K are acceptable as well, and it depends on the domain's specific requirements.} anomalies out of $5,000$ nodes to the domain expert.

A first selection strategy could return the top-30 nodes from a list sorted by the absolute value of the residual (that is the $\delta$). The result is shown in Fig.~\ref{fig:strategy1}. An alternative strategy is to merge two sets to be returned: one set made of the top-15 `negative' outliers, extracted at the head of a list of nodes sorted in ascending order by their residual; another set is made of the top-15 `positive' outliers, extracted at the head of a list of nodes sorted in ascending order by their residual. It's worth noting that the first set is represented by the light orange triangles in the upper part of the chart w.r.t. the identity line, as well as the second set is made of the light orange inverted triangles below the identity line (see Fig~\ref{fig:strategy2}). Although the two strategies produce very similar results in this example (but not identical), they can select very different sets of K anomalies from each other in some other domains. 

\section{Case Study: Anti-Financial Crime Detection in Cross-Country Money Transfer Temporal Networks}\label{sec:casestudy}
Our approach has been applied on a large record of cross-country financial transactions. The ultimate objective is to identify anomalous nodes that, over time, experience unexpected changes in their position within the network. Such nodes warrant further investigation by AFC analysts. Additionally, we can compare different top-K outliers selection strategies to maximize performances. 

\subsection{Dataset description}\label{subsec:dataset}
The dataset supporting the present case study is provided by Intesa Sanpaolo (ISP) and AFC Digital Hub.
It collects all the cross-border transactions that involve ISP's customers or in which ISP’s business is to support the payment performing of partners financial institutions while such operations do not involve its own customers (pass-through). 
It encompasses 80 million SEPA SCT and SWIFT enabled wire transfers in a period of fifteen months, from September 2021 to November 2022. Every transaction has a timestamp indicating day, month and year. We also have the information about both the Country and the BIC code of the sender and the receiver; moreover, for the months of January, February and March 2022, we were also provided with the anonymized IBAN codes involved in the transactions, allowing us to work at multiple levels of temporal and spatial aggregation.

Each dataset row contains the following fields:
\begin{description}
    \item[data\_ref:] transaction date.
    \item[transaction\_id:] internal unique identification code of the transaction.
    \item[BIC:] an anonymised Business Identifier Code (8 digits) of the bank of the payer/beneficiary. 
    \item[IBAN code:] an anonymised version of the International Bank Account Number of the payer/beneficiary (only available for the transactions taking place between January, February and March 2022).
    \item[countryResidence:] ISO code of the country of residence of the payer/beneficiary.
    \item[countryBank:] ISO code of the country of the bank of the payer/beneficiary (according to BIC).
    \item[amount:] expressed in euros. In SEPA transactions and SWIFT transactions where the original currency is euro, the value is actual. In SWIFT transactions where the original currency is not euro, the amount expressed here has been calculated using the change rate at the transaction date, however it may be slightly different from the actual value applied to the client.
    \item[currency:] original currency applied in the transaction.
    \item[source:] data stream of the transaction (SEPA or SWIFT).
\end{description}

The term ``BIC'' denotes the International ISO standard ISO 9362, delineating the structure and components of a universal identifier code named the BIC. This code serves as a requirement for both financial and non-financial institutions, enabling the streamlined automated processing of information on a global scale.
BICs fall into two distinct categories: Connected BICs, which hold access privileges to the Swift network, and non-connected BICs utilized solely for referencing purposes without network access.~\footnote{A BIC code is formatted according to the following structure: https://www.swift.com/standards/data-standards/bic-business-identifier-code, last accessed: 06/02/2024. }
Our dataset includes 8008 unique BICs and 218 countries. The volume of transactions per month is split almost in a 90-10 proportion between SEPA and SWIFT respectively. 

Data were made available to the research team in a fully anonymised form respecting the strictest privacy and security requirements. The data supporting the findings of this study is available from ISP upon request to AFC Digital Hub. Please note that restrictions for data availability apply. Researchers interested in having access to data for academic purposes will be asked to sign a non-disclosure agreement~\footnote{Write to adh@pec.afcdigitalhub.com for further information.}.

Given the information contained in our dataset, we are able to model the transactions at different levels of aggregation, studying money transfers between either countries, BIC codes or IBAN codes. While in Appendix~\ref{app:dataset} we provide a comprehensive data exploration of our dataset, in the next sections we will focus on showcasing the application of our anomaly detection pipeline on these three aggregations.

\subsection{Application to the transactions dataset: identifying anomalous Countries}\label{subsec:countries}

As a first, qualitative step, we aggregate money transfers by countries, in order to identify those nations that, according to our methodology, stand out as potentially anomalous actors in a selected time period. 

The application of the first three steps of the pipeline consists in building the graphs and computing the centrality metrics and node rankings. By considering a monthly temporal resolution, we obtain 15 temporal layers $G_{\mbox{Sep}2021}^\text{Country}, G_{\mbox{Oct}2021}^\text{Country}, \ldots , G_{\mbox{Nov}2022}^\text{Country}$. Nodes represent countries in our dataset. Every temporal layer has its own edge list, such that each link represents that a given amount of money is transferred from one country to the other during that timeframe; the edge will be weighted according to the total amount of money moved, even across multiple transactions, from the source node to the destination. We have a total of 218 countries; if a country does not have transactions in a certain month, it will appear as a singleton in the relative graph. A descriptive overview of the graphs can be found in Tab.~\ref{tab:graphstatscountries}.

\begin{table}[h!]
    \centering
    \resizebox{\linewidth}{!}{%
\begin{tabular}{l|ccccccc}
\toprule
{} &    Edges &  Density &  Avg Degree &  Avg Strength &  Avg Clustering Coeff &  Diameter &  Avg Path Length \\
\midrule
mean  &  4,540.33 &     0.10 &       42.88 &  1.447617e+09 &                  0.68 &      3.47 &             1.90 \\
std   &   144.21 &     0.00 &        1.45 &  2.899440e+08 &                  0.01 &      0.52 &             0.01 \\
min   &  4,309.00 &     0.10 &       40.65 &  1.034154e+09 &                  0.66 &      3.00 &             1.89 \\
max   &  4,750.00 &     0.11 &       45.02 &  1.850451e+09 &                  0.71 &      4.00 &             1.91 \\
\bottomrule
\end{tabular}}

    \caption{Network statistics for the graphs $G^{\text{Country}}$ across the 15 months. Each graph contains 218 countries; if a Country does not have transactions in a certain month, it will appear as a singleton in the relative graph.}
    \label{tab:graphstatscountries}
\end{table}

On each graph, we compute a set of centrality metrics that were described and discussed in Sec.~\ref{subsec:centralities}. Specifically, we will evaluate our nodes based on (in/out)strength, PageRank, hub and authority scores. We are not using (in/out)degree because graphs are weighted and strength can be more informative in this case study. Moreover, we are overlooking other classical centrality measures like closeness and betweenness because these graphs show very short average distances (with very small standard deviation): the nodes would have a very similar closeness with each other, and bridges loses their informational value.

\begin{figure}[h!]
\centering
    \includegraphics[width = \textwidth]{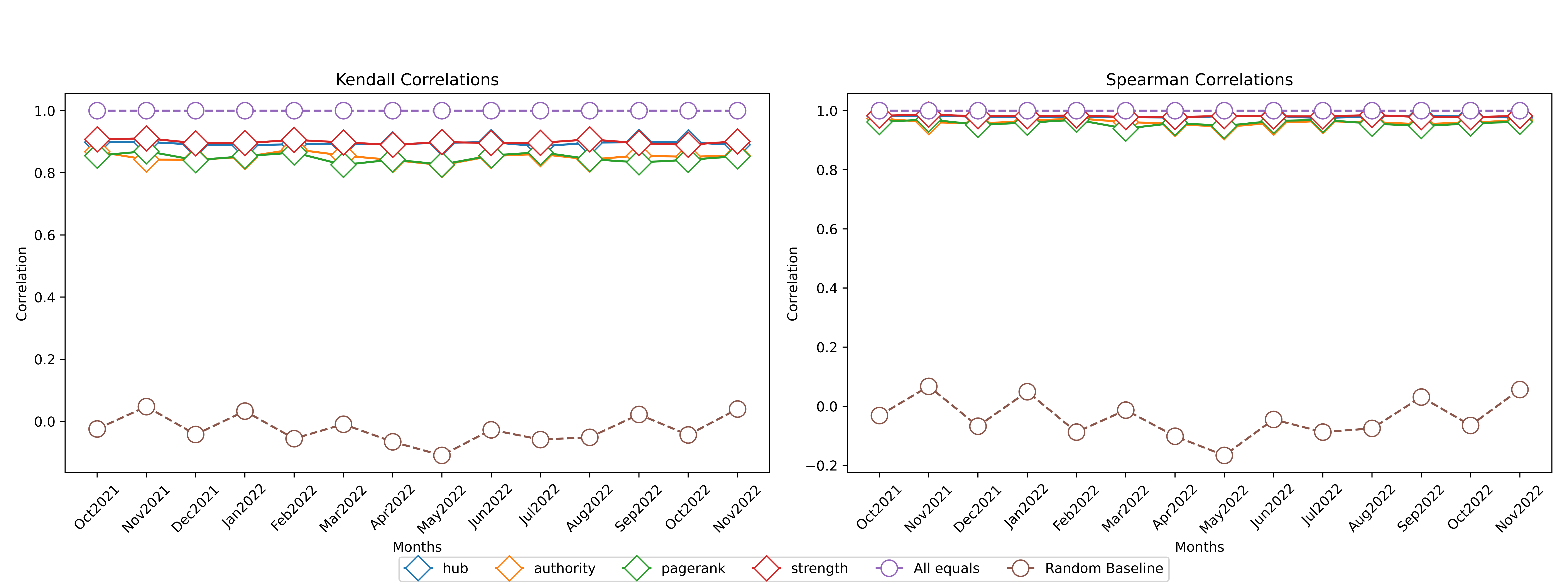}
\caption{Stability over time of the node rankings based on centrality metrics, at the level of countries. The stability in each month is compared to the previous and tested against two baselines: \textit{all equals}, where no change occurs from one ranking to the other, and a \textit{random baselines} where every ranking is randomly reshuffled. The very high levels of correlation, according to both Kendall's and Spearman's coefficients, are consistent over time, suggesting the presence of a condition of stability where nodes tend to preserve their ranking positions.}
\label{fig:ranking_stability_ctry}
\end{figure}

Steps 4 and 5 consist in defining a temporal property and assessing its stability. We observe in Fig.~\ref{fig:ranking_stability_ctry} that the centrality-based node rankings are extremely highly correlated according to both Kendall's $\tau$ and Spearman's $\rho$, and that this is consistently true throughout the entire dataset. According to the formalism introduced in Sec.~\ref{sec:methods}, this is equivalent to setting the temporal property $\overline{P}(G, T_x, T_y, r(m), i)$ to the condition $r_x(m,i) \approx r_y(m,i)$ for any two $(x,y) \in T$ and for any given network centrality $m = \{s^-,s^+, \mbox{pr}, h, a\}$; the evaluation of the rank correlations in Fig.~\ref{fig:ranking_stability_ctry} tells us to which extent this temporal property is satisfied. The rankings based on the centrality metrics are consistently similar over time, and this is checked against two baselines: \textit{All Equals}, where the ranking never changes, resulting in a perfect correlation, and a \textit{random baseline}, where the rankings are randomly shuffled, resulting in a correlation that oscillates around 0.

\begin{figure}[h!]
    \centering
    \includegraphics[width = 0.9\textwidth]{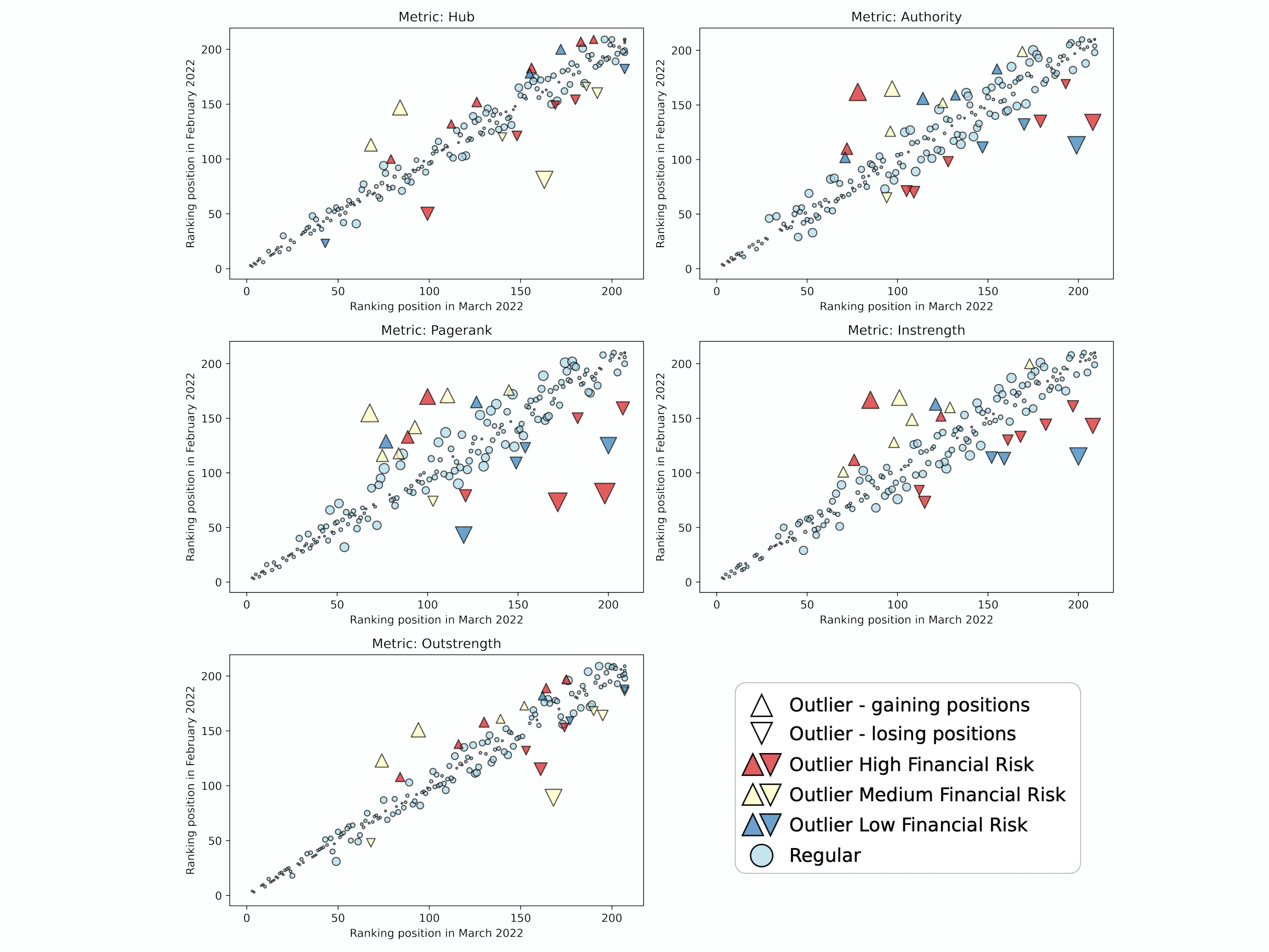}
    \caption{Ranking evolution charts of the time-varying comparison between countries in February/March 2022. The top-10 nodes by residual that are gaining positions (triangles) plus the top-10 that are losing positions (inverted triangles) in the rankings are colored by their financial risk according to ISP internal classification. The rest of the nodes are labeled as ``regular''. Size is proportional to the residual. While there is a clear trend of stable nodes very close to the identity line (as suggested by Fig.~\ref{fig:ranking_stability_ctry}), there is a selection of outliers with high residual that undergo significant changes in their ranking positions.}
    \label{fig:step6_ctry}
\end{figure}

Having passed the necessary check of Step 5, we can proceed with the identification of the top-K anomalies. In step 6 we execute the time-varying comparison between two timeframes, namely February and March 2022~\footnote{Although the method can be applied to any two timeframes, these periods have been selected because of the Russo-Ukrainian War that began in February 2022. This period has been particularly observed by AFC experts, and some more straightforward interpretation and validation of the outcomes was expected}. In Fig.~\ref{fig:step6_ctry} we see the resulting ranking evolution charts, that display how the vast majority of nodes lie very close to the identity line, with only a minority of countries that deviate from the condition of stability identified in steps 4 and 5. The output in each metric is sorted by residual ($\delta$). 

We are now able to evaluate the outliers that emerge from step 6 as potentially anomalous countries. To do so, we refer to the ISP classification of Countries according to their financial risk\footnote{This list is the byproduct of international regulations, company policies and internal information; however, many similar lists are available, such as https://t.ly/Lvwcd~\cite{europeancommission} or https://t.ly/qn2iK~\cite{lawsociety} (last accessed: 24/01/2024).}. This list, which is recurrently updated, classifies countries following a low-medium-high risk scheme. Then, we take the top-10 nodes that are gaining positions plus the top-10 that are losing positions in the rankings to form our top-20 outlier lists. As we can gather from Fig.~\ref{fig:step6_ctry},  many nodes in those outlier lists are high or medium risk countries. Although we are not authorized to disclosed the names and the classification of the individual countries, we can treat high-risk and medium-risk countries as \textit{true positives}, thus quantifying the performances of the algorithm in terms of the average precision-at-k of the lists of outliers generated by each metric. In Tab.~\ref{tab:performance_countries} is reported the average precision-at-k for different values of $k$ for the top-20 anomalies according to every centrality metric. Many centralities are already effective in identifying countries that present medium and high financial risk. This methodology is effectively able to identify nodes that are already known to be, to different extents, potential malicious actors. Interestingly though, not all the reported anomalies are high or medium financial risk countries. While these can be considered as false positives with respect to the financial risk ground-truth, their co-existence in the list of anomalies with ``the usual suspects'' suggests that they are undergoing a highly unexpected change in their role within the network. This information can be extremely valuable in helping AFC analysts navigating such a high load of information, by bringing to their attention potential threats that would not be noticed otherwise, since they are not already on the experts' watch-list. More specifically, since a node can be flagged as anomalous according to more than one metric, repetitions are possible; the union of all the top-20 anomalies based on the five centrality metrics contain $54$ unique countries, divided into $22$ (out of 82) countries with high, $18$ (out of 62) with medium, and $14$ (out of 68) with low financial risk. 
While the limited number of nodes involved allows us to skip step 7, we apply step 8 to obtain a final output. At this stage, the AFC analysts can either choose to evaluate the outliers according to an individual centrality metric, or they can merge one or more lists. Considering all the centralities, we obtain a list of $54$ countries that can be sorted according to an external criteria, in the hope of shifting more significant nodes towards the first positions of the final list. Under the guidance of the domain experts we sorted the output based on the node's change in volume of money exchanged with countries with high financial risk with respect to the previous timeframe (\textit{delta amount high risk} $\Delta_{HRA}$), as this can be an interesting marker of potential fraudulent behaviour.
The performances are evaluated in Tab.~\ref{tab:step8_ctry}; this strategy achieves good performances in terms of both precision-at-k and recall. 

The output of this pipeline provides the user with useful information: as argued, it is able to identify among the vast amount of transactions data a number of well known malicious agents, together - and, maybe, most importantly - with other potentially anomalous nodes that are more unexpected. Anyway, such a coarse-grained analysis, even though yielding very encouraging results, is still preliminary. We can fully exploit the potential of this methodology by examining transactions at a finer grain, identifying outliers that can be thoroughly inspected by AFC experts in order to provide more precise information at a smaller scale.

\begin{table}[]
    \centering
    \begin{tabular}{lccccccccc}
\toprule
{} &  p@1 &  p@2 &   p@5 &  avg p@5 &  avg p@10 &  avg p@20 &  r at 20 \\
\midrule
hub         &  1.0 &       1.0 &        1.0 &     1.00 &      0.95 &      0.88 &     0.80 \\
authority   &  0.0 &       0.5 &        0.8 &     0.68 &      0.68 &      0.66 &     0.60 \\
pagerank    &  1.0 &       1.0 &        0.6 &     1.00 &      0.82 &      0.77 &     0.70 \\
instrength  &  0.0 &       0.5 &        0.6 &     0.64 &      0.63 &      0.70 &     0.80 \\
outstrength &  1.0 &       1.0 &        1.0 &     1.00 &      1.00 &      0.99 &     0.85 \\
\bottomrule
\end{tabular}
\caption{Performances of the algorithm in terms of the average precision-at-k and recall of the lists of outliers generated by each metric. At this stage of analysis, since we have the full ground-truth of the financial risk, we are able to compute the recall score.}\label{tab:performance_countries}
\end{table}

\begin{table}[]
    \centering
    \resizebox{\linewidth}{!}{
    \begin{tabular}{lcccccccc}
\toprule
{} &  p@1  &  p@2 &  p@5 &  avg p@5 &  avg p@10 &  avg p@20 &  avg p@30 &  r at 30 \\
\midrule
Mixed sorting strategy by $\Delta_{HRA}$ &  1.0 &   1.0 &  1.0 &      1.0 &       1.0 &      0.93 &      0.86 &     0.73 \\
\bottomrule
\end{tabular}}
    \caption{Evaluation of step 8: performance obtained by merging the output of step 6 in a final, single list of outliers sorted by $\Delta_{HRA}$. }
    \label{tab:step8_ctry}
\end{table}

\subsection{Application to the transactions dataset: identifying anomalous BIC codes}\label{subsec:bic}

We now aggregate the transactions at the finer level of BICs and we apply our outlier detection pipeline in order to identify potentially anomalous BIC codes in our data. By considering the same monthly temporal resolution, we obtain 15 temporal layers $G_{\mbox{Sep}2021}^{\mbox{BIC}}, G_{\mbox{Oct}2021}^{\mbox{BIC}}, \ldots, G_{\mbox{Nov}2022}^{\mbox{BIC}}$. Every graph has 8008 nodes, and each node represents a different (anonymized) BIC code, namely, the branch of a bank in a certain Country. In each layer, an edge between two nodes is established if, at some point within the considered time window, money is transferred from one BIC to the other; the edge will be weighted according to the total amount of money moved, even across multiple transactions, from the source node to the destination. BICs that, during that month, do not have exchange money with other BICs, will be singletons. A descriptive overview of the graphs can be found in Tab.~\ref{tab:graphstatsbics}.

\begin{table}[h!]
    \centering
    \resizebox{\linewidth}{!}{%
\begin{tabular}{l|ccccccc}
\toprule
{} &     Edges &  Density &  Avg Degree &  Avg Strength &  Avg Clustering Coeff &  Diameter &  Avg Path Length \\
\midrule
mean  &  58,130.73 &      $1.8 10^{-3}$ &       20.48 &   54,024,165.98 &                  0.28 &      6.20 &             2.78 \\
std   &   1,734.11 &      $1.0 10^{-4}$ &        0.66 &   10,903,596.64 &                  0.00 &      0.41 &             0.02 \\
min  &  55,509.00 &     $1.7 10^{-3}$ &       19.47 &   38,442,552.78 &                  0.27 &      6.00 &             2.74 \\
max  &  61,142.00 &      $1.9 10^{-3}$ &       21.69 &   68,983,766.61 &                  0.29 &      7.00 &             2.82 \\
\bottomrule
\end{tabular}}

    \caption{Network statistics for the graphs $G^\mathcal{BIC}$ across the 15 months. Each graph contains 8008 BICs.}
    \label{tab:graphstatsbics}
\end{table}

\begin{figure}[h!]
\centering
    \includegraphics[width =\textwidth]{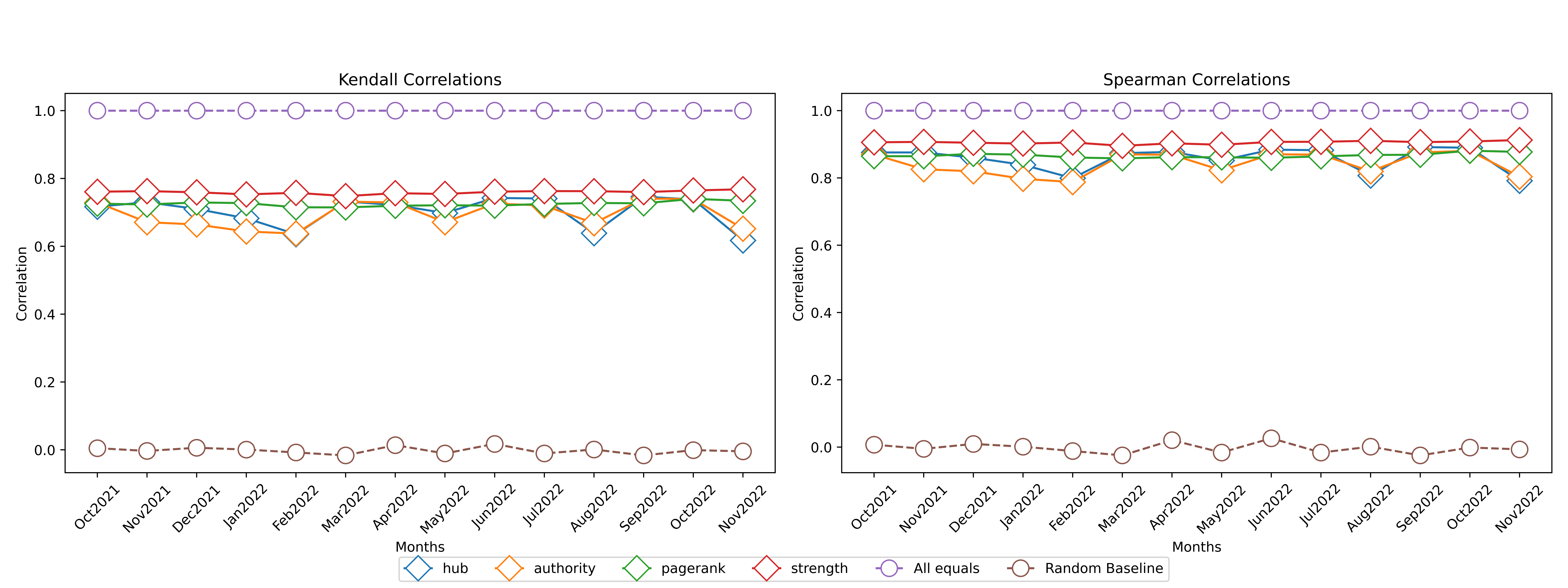}
\caption{Stability over time of the node rankings based on centrality metrics. The stability in each month is compared to the previous and tested against two baselines: \textit{all equals}, where no change occurs from one ranking to the other, and a \textit{random baselines} where every ranking is randomly reshuffled. The high levels of correlation, according to both Kendall's and Spearman's coefficients, are consistent over time, suggesting the presence of a condition of stability where nodes tend to preserve their ranking positions.}
\label{fig:ranking_stability}
\end{figure}

As done for the Country graphs, we build node rankings based on the same network metrics (i.e., (in/out)strength, PageRank, hub and authority scores) and we apply step 4 and 5 of the pipeline to identify a temporal property and to assess its stability over time. Fig.~\ref{fig:ranking_stability} suggests that there is a consistent stability of the node rankings over time for all the centrality measures we selected. This stability is measured according to Spearman's $\rho$ and Kendall's $\tau$.




\begin{figure}
    \centering
    \includegraphics[width = .7\textwidth]{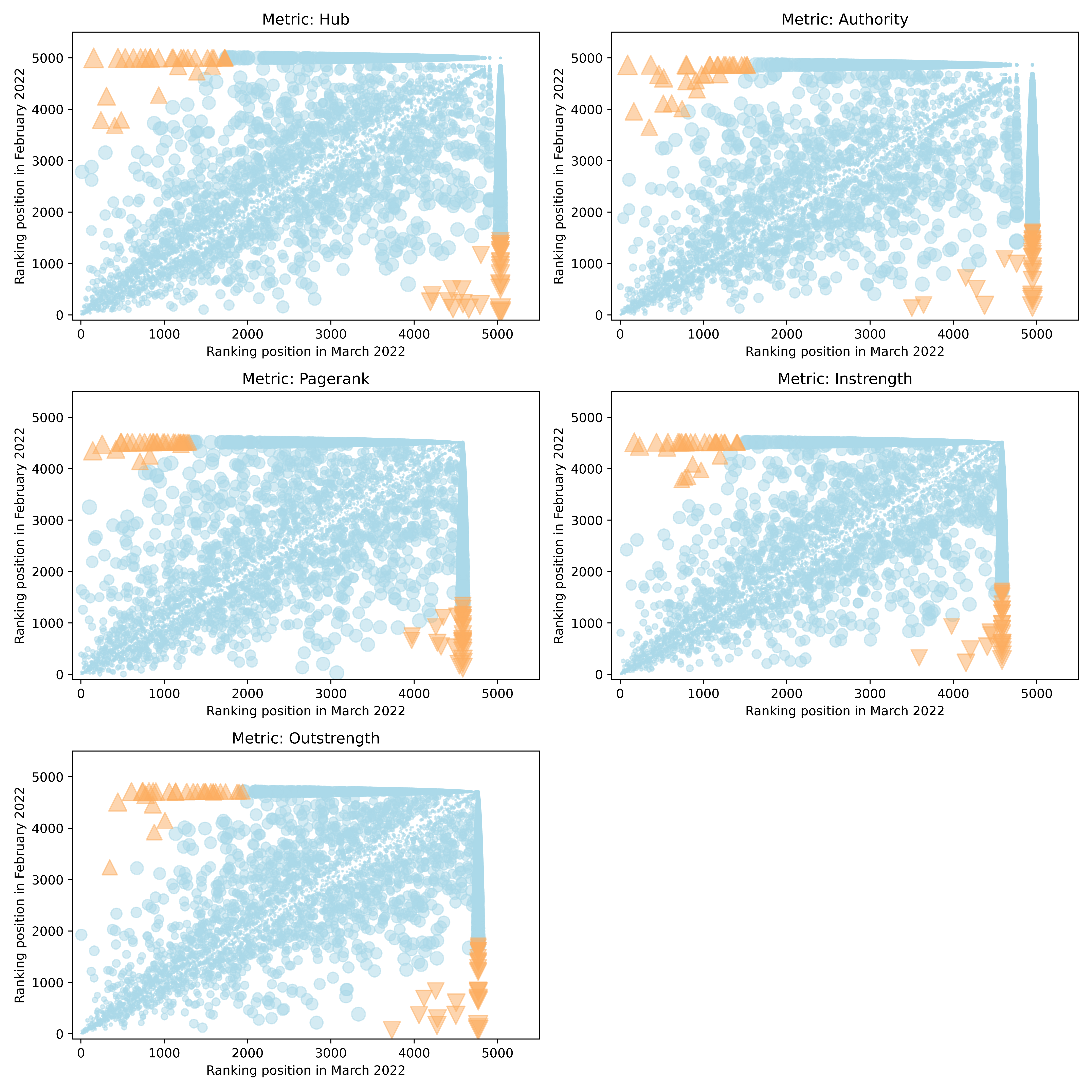}
    \caption{Step 6 of the pipeline applied on the BIC networks. The data points in light orange are the top-30 nodes that are gaining positions (triangles) plus the top-30 that are losing positions (inverted triangles) in the rankings from Feb 2022 to Mar 2022. Size is proportional to the absolute value of the residual. While there is a clear trend of stabler nodes close to the identity line (as already shown in Fig.~\ref{fig:ranking_stability}), there is a selection of outliers with high residual that undergo significant changes in their ranking positions from February to March 2022.}
    \label{fig:step6}
\end{figure}

Having successfully passed the stability check of step 5, we proceed with the identification of the top-K anomalies according to every network centrality metric.
We take into consideration the same timeframes of February and March 2022.
In Fig.~\ref{fig:step6} we can evaluate the resulting \textit{ranking evolution charts}. The data points in orange are the top-30 nodes by residual that are gaining positions plus the top-30 that are losing positions in the rankings from Feb 2022 to Mar 2022. The cutoff of 60 nodes per metric was chosen based on experimental needs; as argued in Sec.~\ref{subsec:scope}, the analyst needs a \textit{reasonable} number of anomalies to inspect manually due to time and resources constraints. The appropriate number therefore strictly depends on the specifics of the annotation process and, in this specific case, was suggested by the domain experts.

Considering the significantly larger quantity of data points compared to the prior study concerning transactions aggregated by countries, we added here an optional filtering phase (step 7) to prevent overloading the AFC analyst nodes with data that fails to satisfy certain baseline requirements. We employed a \textbf{threshold based filter} in this process. Adhering to the guidelines set by domain experts, we applied filters to the total strength, which refers to the total sum of money transacted by the node, at the BIC level of aggregation. More precisely, we keep nodes that correspond to:
\begin{itemize}
    \item BICs belonging to Countries with \textbf{low} and \textbf{medium} financial risk - according to the ISP internal classification - that move more than $5 \cdot T_{hr}$ Euros in at least one of the two timeframes;
    \item BICs belonging to Countries with \textbf{high} financial risk - according to the ISP internal classification - that move more than $T_{hr}$ Euros in at least one of the two timeframes
\end{itemize}

where $T_{hr}$ is an internal threshold of attention for high financial risk countries that will remain undisclosed. This procedure removes 29 unique BICs across the different centrality metrics.

\begin{figure}[h!]
\centering
    \begin{subfigure}[b]{0.49\textwidth}
        \centering
        \includegraphics[width=0.9\textwidth]{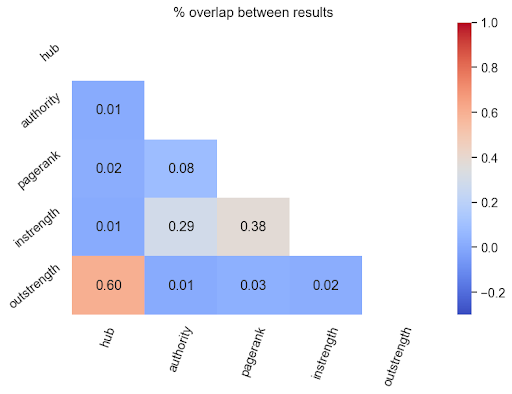}
        \caption{} 
        \label{subfig:cor1}
    \end{subfigure}
    \begin{subfigure}[b]{0.49\textwidth}
        \centering
        \includegraphics[width=0.9\textwidth]{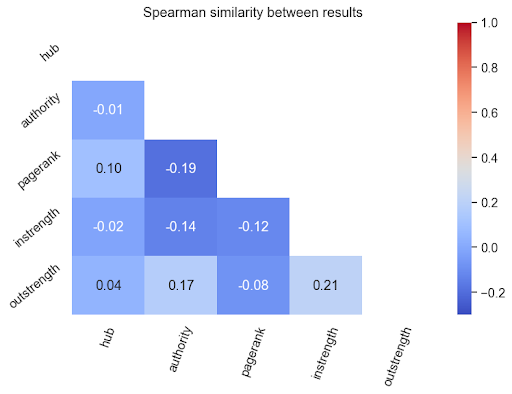}
        \caption{} 
        \label{subfig:cor2}
    \end{subfigure}
\caption{Correlations between the outputs on the different metrics. The similarity of the outputs in the case of BICs is slightly lower compared to the Country level of analysis. Nonetheless, the correlation between the outputs on the different metrics indicates that all centrality measures should be taken into account while forming the final list to avoid losing too much information.}
\label{fig:correlation_output}
\end{figure}

After generating the filtered outliers list for each centrality metric, our observations reveal that, as evidenced in Fig.~\ref{fig:correlation_output}, there is a notable correlation between the rankings produced by  outstrength and hub score, along with a weaker yet still positive correlation between those produced by instrength and both PageRank and authority scores. The Spearman correlation among these rankings is quite low, indicating the value in considering all outputs.

Therefore, we advance to the final step 8. The comparative analysis, which will be further discussed in Sec.~\ref{subsec:annotation}, corroborates that no single strategy distinctly surpasses others. This necessitates a hybrid approach that amalgamates the strengths of all five top-60 lists, leading to the formation of a comprehensive, final list of outliers that integrates insights from various centrality metrics. As a fundamental baseline, we employ the same \textbf{mixed sorting strategy} used in the preceding section: we compile the union of all top-60 nodes and then rank them according to the $\Delta_{HRA}$ metric. In addition, we have also developed a \textbf{stratified sorting strategy}, which is executed as follows:

\begin{enumerate}
\item we group all nodes that share identical positions across all the top-K lists;
\item we sort each group internally by $\Delta_{HRA}$, i.e., according to the node's change in volume of money exchanged with countries with high financial risk with respect to the previous timeframe;
\item we remove the duplicated entries by keeping the highest one, as one node can be identified as an outlier according to more than one centrality metric.
\end{enumerate}

Thus, the stratified approach preserves the information of the original ranking positions of the outliers, exploiting also the information coming from  $\Delta_{HRA}$.

\subsubsection{Annotation process and evaluation of the results}\label{subsec:annotation}

Given the much more specific and fine-grained quality of the results at this level of aggregation, the domain experts were able to provide a substantial feedback by undertaking a process of manual validation of the results. The goals of this process are:
\begin{itemize}
    \item to assess the proportion of real anomalies (\textit{true positives} within the outliers identified by the pipeline);   
    \item to assess the goodness of the individual network metrics, by evaluating the position of the true positives in every centrality-based ranking;
    \item to assess the goodness of the outliers' sorting strategies by evaluating if the true positives occupy the highest positions of the corresponding final ranked list.
\end{itemize}


Step 6 produces, by comparing the node rankings based on every network centralities, five lists of top-60 outliers sorted by their residual $\delta$. 
We can quantify the performances of each of these rankings through the precision-at-k and the average precision-at-k shown in 
Table~\ref{tab:step6}: lists created after instrength centrality and authority score changes over time are slightly better than the others. A similar evaluation is done after the execution of step 7, which simply consists in applying the threshold-based filter defined in Sec.~\ref{subsec:bic} to remove uninteresting nodes.
In Table~\ref{tab:step7}, we see that the application of this filter slightly improves the performances of some of the rankings: this happens because, by removing nodes, true positive can move upwards in the lists, improving the precision-at-k for the low values of $k$ in which we are interested. The improvement seen in Table~\ref{tab:step7} is only marginal, but no true positive is removed by the filter, suggesting that the algorithm is already effective in identifying interesting nodes even if it has no prior domain knowledge. We tested other ML-based filtering methods at this stage (i.e., Isolation Forest and OneClassSVM), without gaining any improvements, so we skip the discussion of these experiments from the paper. For whom is interested, we describe some results in Appendix~\ref{app:ML}.


\begin{table}[]
    \centering
    \resizebox{\linewidth}{!}{
    \begin{tabular}{lcccccccccc}
\toprule
{} &  p@1 &  p@2 &   p@5 &  avg p@5 &  avg p@10 &  avg p@20 &  avg p@30 &  avg p@60 &  r* at 30 &  r* at 60 \\
\midrule
Hub         &  0.0 &  0.0 &  0.20 &     0.20 &      0.22 &      0.23 &      0.22 &      0.19 &      0.19 &      0.29 \\
Authority   &  1.0 &  1.0 &  0.40 &     1.00 &      0.83 &      0.83 &      0.66 &      0.55 &      0.19 &      0.24 \\
PageRank    &  0.0 &  0.0 &  0.00 &     0.00 &      0.19 &      0.19 &      0.19 &      0.17 &      0.19 &      0.33 \\
Instrength  &  1.0 &  1.0 &  0.40 &     1.00 &      1.00 &      0.76 &      0.76 &      0.39 &      0.14 &      0.33 \\
Outstrength &  0.0 &  0.0 &  0.20 &     0.20 &      0.26 &      0.29 &      0.29 &      0.26 &      0.19 &      0.24 \\
\hline \textbf{Average}     &  0.4 &  0.4 &  0.24 &     0.48 &      0.50 &      0.46 &      0.42 &      0.31 &      0.18 &      0.29 \\
\bottomrule
\end{tabular}}
    \caption{Evaluation of step 6: Time-varying comparison produces, for each metric, top-30 lists of outliers, that are evaluated after the experts' annotation: BICs are labeled as relevant (true positive) or not relevant (false positive), and this is used to calculate performances of the different basic strategies. }
    \label{tab:step6}
\end{table}

\begin{table}[]
    \centering
    \resizebox{\linewidth}{!}{
    \begin{tabular}{lcccccccccc}
\toprule
{} &  p@1 &  p@2 &   p@5 &  avg p@5 &  avg p@10 &  avg p@20 &  avg p@30 &  avg p@60 &  r* at 30 &  r* at 60 \\
\midrule
Hub         &  0.0 &  0.0 &  0.20 &     0.20 &      0.22 &      0.23 &      0.22 &      0.19 &      0.19 &      0.29 \\
Authority   &  1.0 &  1.0 &  0.40 &     1.00 &      0.83 &      0.83 &      0.67 &      0.56 &      0.19 &      0.24 \\
PageRank    &  0.0 &  0.0 &  0.20 &     0.20 &      0.22 &      0.22 &      0.22 &      0.19 &      0.19 &      0.33 \\
Instrength  &  1.0 &  1.0 &  0.40 &     1.00 &      1.00 &      0.76 &      0.76 &      0.39 &      0.14 &      0.33 \\
Outstrength &  0.0 &  0.0 &  0.20 &     0.20 &      0.26 &      0.29 &      0.29 &      0.26 &      0.19 &      0.24 \\
\hline \textbf{Average}     &  0.4 &  0.4 &  0.28 &     0.52 &      0.51 &      0.47 &      0.43 &      0.32 &      0.18 &      0.29 \\
\bottomrule
\end{tabular}}

    \caption{Evaluation of step 7: The application of the threshold based filter slightly improves the basic rankings' performances.}
    \label{tab:step7}
\end{table}

\begin{figure}
    \centering
    \includegraphics[width = 0.95\textwidth]{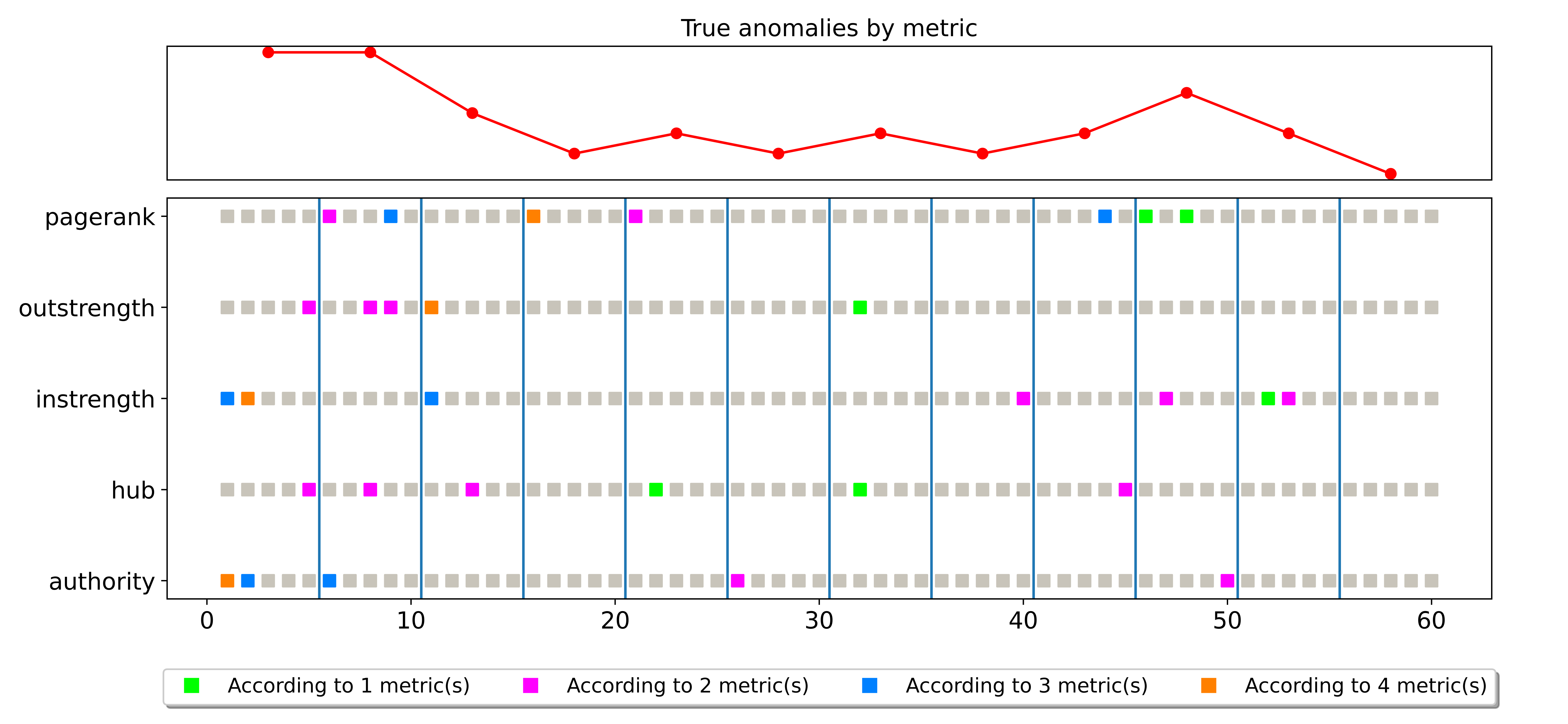}
    \caption{Positions of true positives in the rankings based on the different centrality metrics. The line chart in the top panel represents the ratio true/false positives across all the metrics, showing a decreasing trend as we move away from the top positions of the rankings.} 
    \label{fig:grid_metric}
\end{figure}

In Fig.~\ref{fig:grid_metric}, we assess the placement of true positives within each list. Given that nodes may feature in the outputs of multiple metrics, occurrences of repetition are possible. Nevertheless, it is observable that novel outliers tend to be positioned more prominently at the top of each ranking. This observation advocates for the implementation of a mixed strategy at step 8, which prioritizes nodes already ranking high in the basic listings.

\begin{figure}[h!]
    \centering
    \includegraphics[width = 0.8\textwidth]{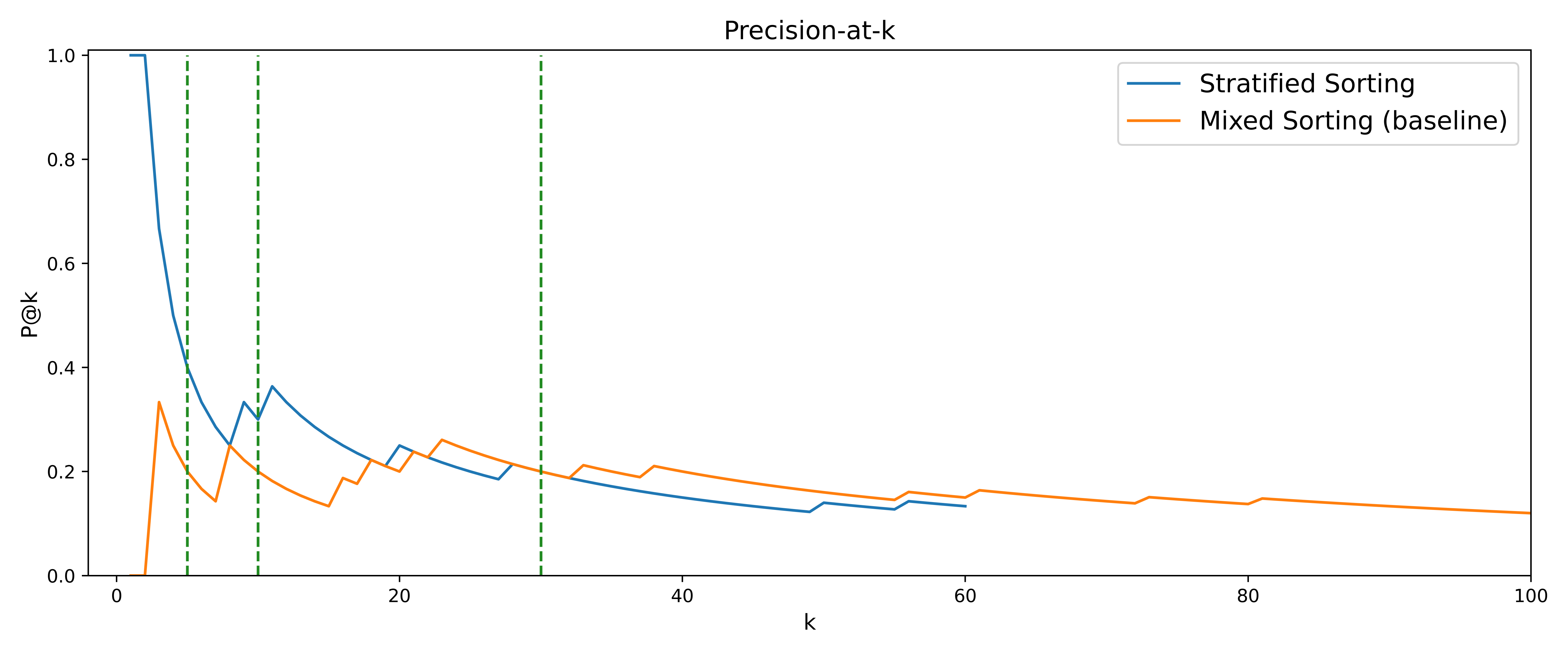}
    \caption{Evaluation of different sorting methodologies in terms of \textit{precision-at-k}. The stratified sorting strategy is compared to a mixed sorting strategy that is used as a baseline.  The stratified strategy is clearly superior, in terms of P@K, in the first 20 positions of the final output.}
    \label{fig:stratified_pak}
\end{figure}

Hence, we adopt the above mentioned stratified strategy, in order to obtain a final, single list of top-60 outliers that embeds the knowledge coming from more than one centrality metric. This is evaluated in Fig.~\ref{fig:stratified_pak} as well as in Tab.~\ref{tab:step8} and tested against the simpler mixed strategy that ranks nodes from the five lists of outliers by their $\Delta_{HRA}$, disregarding the information of their position in the respective rankings that is instead preserved by the stratified sorting strategy. The latter is clearly superior, in terms of P@K, in the first 20 positions of the final output. The average precision-at-10 is 0.78, suggesting a very good performance in the top-10 ranking positions, surpassing also all the other basic ranking strategies based on a single centrality measure.

\begin{table}[]
    \centering 
\resizebox{\linewidth}{!}{
\begin{tabular}{lcccccccccc}
\toprule
{} &  p@1 &  p@2 &  p@5 &  avg p@5 &  avg p@10 &  avg p@20 &  avg p@30 &  avg p@60 &  r* at 30 &  r* at 60 \\
\midrule
Mixed Sorting (baseline) &  0.0 &  0.0 &  0.2 &     0.33 &      0.29 &      0.25 &      0.25 &      0.23 &      0.29 &      0.43 \\
\textbf{Stratified Sorting}       &  1.0 &  1.0 &  0.4 &     1.00 &      0.78 &      0.59 &      0.53 &      0.43 &      0.29 &      0.38 \\
\bottomrule
\end{tabular}}
\caption{Evaluation of Step 8: We compare the performances of a mixed sorting baseline versus the stratified strategy, that outperforms the first.}
\label{tab:step8}
\end{table}

\subsection{Application to the transactions dataset: going deeper with IBAN codes}\label{subsec:IBAN}

We can apply our anomaly detection pipeline to an even finer-grained level, trying to identify anomalies among individual bank accounts. Indeed, the dataset provided contains, for every transaction between January and March 2022, information about the anonymized IBAN codes involved as sender and receiver of every wire transfer. 

By considering the same monthly temporal resolution, we obtain 3 temporal layers $G_{\mbox{Jan}2022}^{\mbox{IBAN}}, G_{\mbox{Feb}2022}^{\mbox{IBAN}}$ and $G_{\mbox{Mar}2022}^{\mbox{IBAN}}$. In each graph, nodes represent bank accounts (originally identified by their IBAN codes) that will be linked by an edge if, at some point within the considered time window, money is transferred from one account to the other; the edge will be weighted according to the total amount of money moved, even across multiple transactions, from the source node to the destination. We obtain extremely sparse graphs ($\sim 3M$ nodes and $\sim 3M$ edges, for a density $\sim 10^{-7}$).

On each graph we apply steps 2 and 3 of the pipeline, building the centrality-based node rankings, and we look for the validity and stability of the temporal property through steps 4 and 5. Unfortunately, 
unlike the case of BICs, there is a constantly low correlation ($< 0.5$) between each ranking, not enough to ensure the required stability. Therefore, since the stability check fails, we are not able to apply the rest of the pipeline of anomaly detection by looking for outliers with respect to their variations of ranking positions.

Nonetheless, even though we cannot apply the anomaly detection pipeline at this scale, the outputs of the previous Section on the BIC analysis can be used as a starting point to explore the transactions between IBANs through other techniques of network analysis.

   
   

This is not entirely unexpected. As stated, when building the graph we obtain very sparse networks. Quoting a seminal study on sparse graphs~\cite{bollobas2011sparse}: ``[...] graphs with $\Theta(n)$ edges (...) turn out to be much harder to handle. Many new phenomena occur, and there are a host of plausible metrics to consider [...]''. Even though there are visible differences between regular BICs and outliers, as we can see in Appendix~\ref{app:iban}, suggesting that something is going on and could be detected also at the level of IBANs, causality is not easy to establish. This implies that we are not able to apply the same ideas on anomaly detection considered for the BIC graphs exactly as they are. Nonetheless, many strategies could be applied to exploit the information obtained at the level of Countries and BICs. For instance, once that the anomalous BICs are identified as an output of the pipeline, it is possible to investigate such anomalies by expanding the networks of IBANs underlying the anomalous BICs. In this way, the AFC analyst is able to exploit network-based techniques to gain precious insights on the fine-grained transactions between individual accounts, digging into the exchanges of money that generated the anomaly at the BIC level. An example of this approach can be seen in Appendix~\ref{app:iban}.

\begin{figure*}[ht!]
\centering
    \begin{subfigure}[b]{0.49\textwidth}
        \centering
        \includegraphics[width=0.9\textwidth]{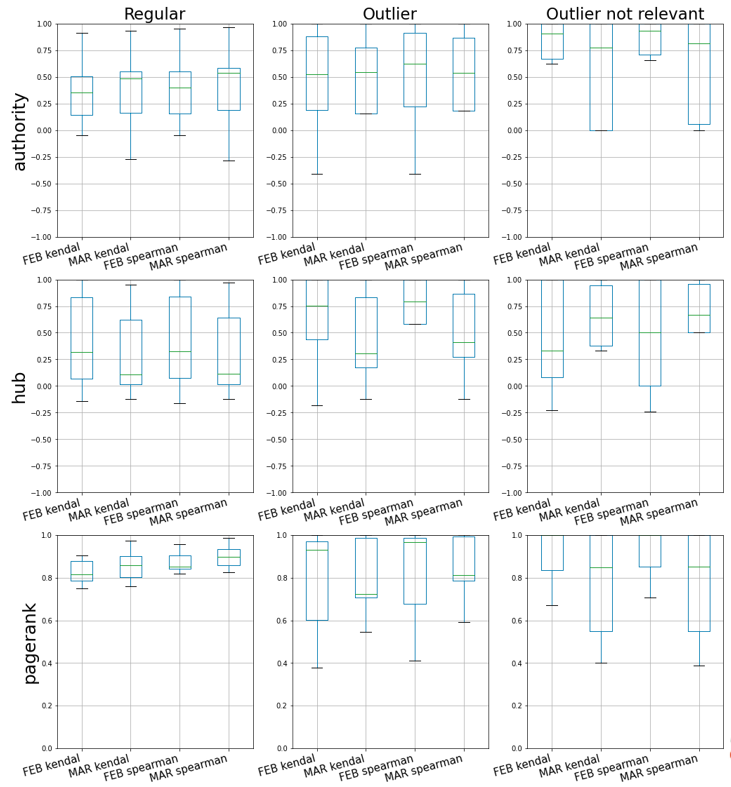}
        \caption{}
        \label{fig:metrics_box}
    \end{subfigure}
    \begin{subfigure}[b]{0.49\textwidth}
        \centering
        \includegraphics[width=0.9\textwidth]{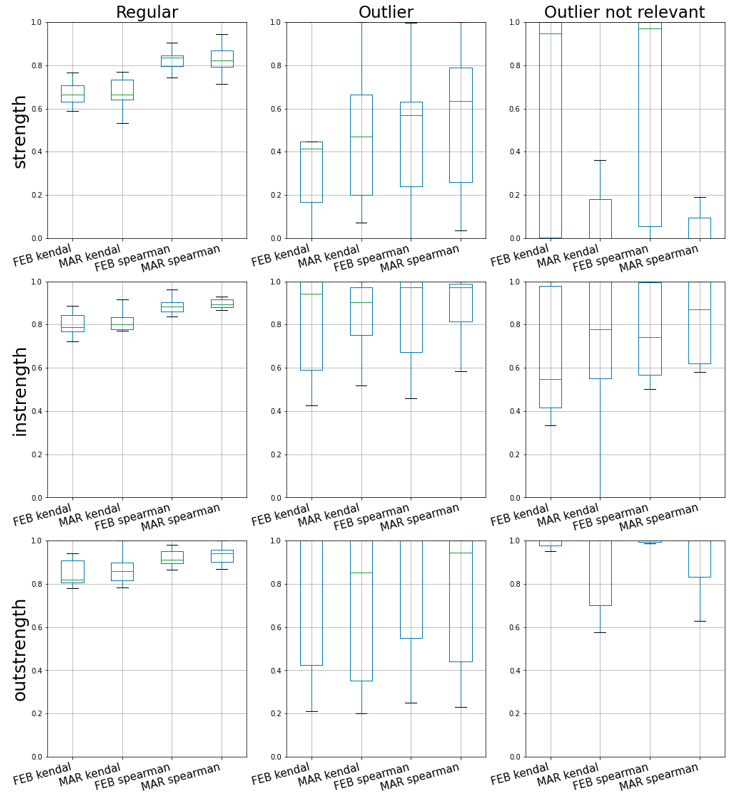}
        \caption{} 
        \label{fig:metrics_box2}
    \end{subfigure}
    \caption{Distribution of the correlation for node rankings generated on different network metrics. Specifically, Spearman and Kendall correlations are computed for the rankings of a month with respect to the previous. This is done on the IBAN networks underlying ``regular'' BICs, `` outlier'' BICs (i.e., true positives) and ``outliers not relevant '' (i.e., false positives). We notice how larger is the distribution in the latter cases with respect to regular BICs.} \label{fig:metrics_boxplot}
\end{figure*}

\subsection{Discussion}

The anomaly detection pipeline defined in this work leverages the observation of the evolution of node rankings over time to identify potential anomalies in the system. It was built around the assumptions and requirements identified in Sec.~\ref{subsec:scope}. In this regard, our methodology addresses the problem by:

\begin{itemize}
    \item producing, as a final output of the algorithm, a sorted list that maximises the probability of finding true positives at the top positions of the final output. As outlined in assumption \textbf{A1} and in requirement \textbf{R2}, the \textit{human-in-the-loop} process of verifying anomalies is time-consuming, and the ideal expert system would provide the analyst with a concise solution, providing a small set of nodes that are very likely to be actual anomalies;
    \item introducing exogenous information in the decisional process. As stated in assumption \textbf{A2}, even the expert cannot confirm whether the node identified as an actual anomaly corresponds to an actual violation, until the appropriate information is examined by the competent authority. In order to reduce the chance of false positives, it is important to incorporate the knowledge of the domain expert in the algorithm;
    \item \textbf{O3}: finally, as outlined in assumption \textbf{A3}, a full interpretability of the reported anomalies is required by regulators. While leveraging the embedded complexity of the financial network, our approach ensures transparency and clarity in the interpretation of results thanks to the clear interpretation of centrality metrics. This aligns with the need for regulatory actions to have clear rationales behind them.
\end{itemize}

This approach can be applied to any system that can be modeled through networks, proven that certain conditions are met. The use case proposed in the financial domain was particularly illustrative since, given the peculiarity of the data, we were able to apply the same process at different levels of aggregation, showcasing the potential of this methodology. We explored the dataset provided by Intesa Sanpaolo, looking for potential anomalies in the transactions between Countries, BICs and IBANs. The experimental pipeline, detailed in Sec.\ref{subsec:pipeline} and in Fig.~\ref{fig:scheme}, is organized in steps: starting from the creation of the graphs, it then proceeds in calculating the node rankings based on a set of network centrality metrics, observing their evolution and keeping track of the role of the nodes over time. It finds its cornerstone in the examination of temporal stability across the various centrality metrics: the observed consistency in node rankings, serves to identify the needed condition of stability out of which the outliers are defined. This requirement is not always satisfied, as seen in Sec.~\ref{subsec:IBAN}: whenever the temporal stability of the centrality-based rankings is not satisfied, it is not possible to identify outliers based on their deviation from the norm. 

This was not the case for BICs and Countries, where the consistent stability enabled us to proceed with the time-varying comparisons. The anomalies resulting from the application of the subsequent steps of the pipeline on the Country graphs constitute, qualitatively, interesting results, since many of the top-20 anomalous nodes are nations that are already known to be involved in potentially malicious financial activities. While is not possible to disclose any details about real cases linked with the operational activity of Intesa Sanpaolo on which the model has been applied, it is possible to make reference to publicly known cases that totally fit its approach and show its empowering vale in transaction monitoring against financial crime. At the time being, several media sources\footnote{https://www.reuters.com/world/german-exports-russias-neighbours-fuel-sanctions-evasion-fears-2023-05-16/.} have reported the sharp decrease of German exports to Russia in the first quarter of 2023 due to EU financial sanctions against this country. Such discontinuity is almost contemporary to the unprecedented increase of the same export flow to Kyrgyzstan arising evident but heavily delayed suspicion about circumvention of sanctions practices. Specifically, the value of German exports to Russia itself slumped by more than 47\% in January-March compared with the same period a year earlier, reflecting tough restrictions on trade imposed by the European Union and other Western powers. However, exports from Germany to Kyrgyzstan rose some 949\%, to 170 million euros (187.14 million dollars), a Reuters analysis based on data from the German statistics office shows.
Due to the fragmentations of such exporting flows among several chains of players as well as the not direct bordering of the two countries, the inferred circumventing triangulations between Germany and Russia by Kyrgyzstan has been gone undetected for months.
Indeed, the adoption of single transactions or groups of transactions linked to specific accounts as unit of analysis, may jeopardize the ability of a transaction monitoring system to spot, inside the relevant level of noise physiologically affecting the financial flows, the signal of this kind of disrupting shift.
On the contrary, the same phenomenon is apparent adopting a top-down, macro-level, network-based methodology such as the one currently described.

The validation carried out on the outcomes of the application at the level of BICs instead allowed us to validate quantitatively the different steps of the pipeline. Of the unique nodes handed to the experts following all the steps of the pipeline, 14 are confirmed as true positives worth of further investigations by the analysts and the competent authorities. The final output performs well with respect to the requirement \textbf{R1} defined in Sec.~\ref{subsec:scope}, with good levels of precision-at-k for the top positions of the rankings and an average precision-at-10 of 0.78. 


\section{Conclusions}\label{sec:conclusions}

In this work, we proposed a novel methodology to perform anomaly detection on complex networks. It leverages the insights provided by network analysis and centrality measures to produce a ranked output of potential anomalies, suggesting a set of nodes that undergo unexpected changes in their role in the network. This approach was designed based on the assumptions and the requirements explained in Sec.~\ref{subsec:scope}: it entirely revolves around simplifying the life of the domain expert, who will need to minimise the time spent navigating through the large amount of complex data while looking for potential anomalies.

This methodology aims at filtering out the majority of data points that follow what is identified as a normal behaviour, leaving only a reduced number of items to the attention of the expert, and it is suitable to be applied in the specific context when there are no expectations whatsoever about the number of anomalies, nor knowledge about specific behaviour that should be looked for. For this reason, a good choice of network metrics - combined with a precise ranking strategy and, whenever possible, with quantitative domain knowledge, is crucial in ensuring that, even if an arbitrary small choice is made in terms of output size, the probability of finding real anomalies is maximised. Furthermore, every node identified as a potential anomaly has been included in the final output due to its unexpected behaviour with respect to one or more node centrality metrics, ensuring a full interpretability of the result. To the best of our knowledge, the methodology is novel and no similar approach has yet been tested in AFC practices. Although it is based on quite simple temporal observations of ranking produced by basic centrality measures, it constitutes a radical shift from the current AFC methodologies, providing the AFC analysts with a top-down view of the system, enabling them to uncover potential threats that could have gone unnoticed otherwise, or could have required a substantially higher effort. 

While this method has been developed and tested within the financial domain, it can be easily applied to other contexts. Potentially, every system that can be modeled through graphs and that evolves in time can be studied with this approach. Potential extensions of this methodology could involve applying it and, eventually, tailoring it to other domains, such us biology and genomics, or social networks. While we leave this as future work, we believe that these results constitute a positive contribution in improving the state of the art on network analysis applied to the detection of anomalies in the financial domain, helping AFC analysts in a delicate context where, due to an intricate tangle of contextual and regulatory reasons, the firepower of deep learning-based methods sometimes cannot be deployed.

\newpage

\section*{Appendix}
\appendix
\section{Dataset Statistics}\label{app:dataset}

\begin{figure}[ht!]
\centering
\includegraphics[width=0.8\textwidth]{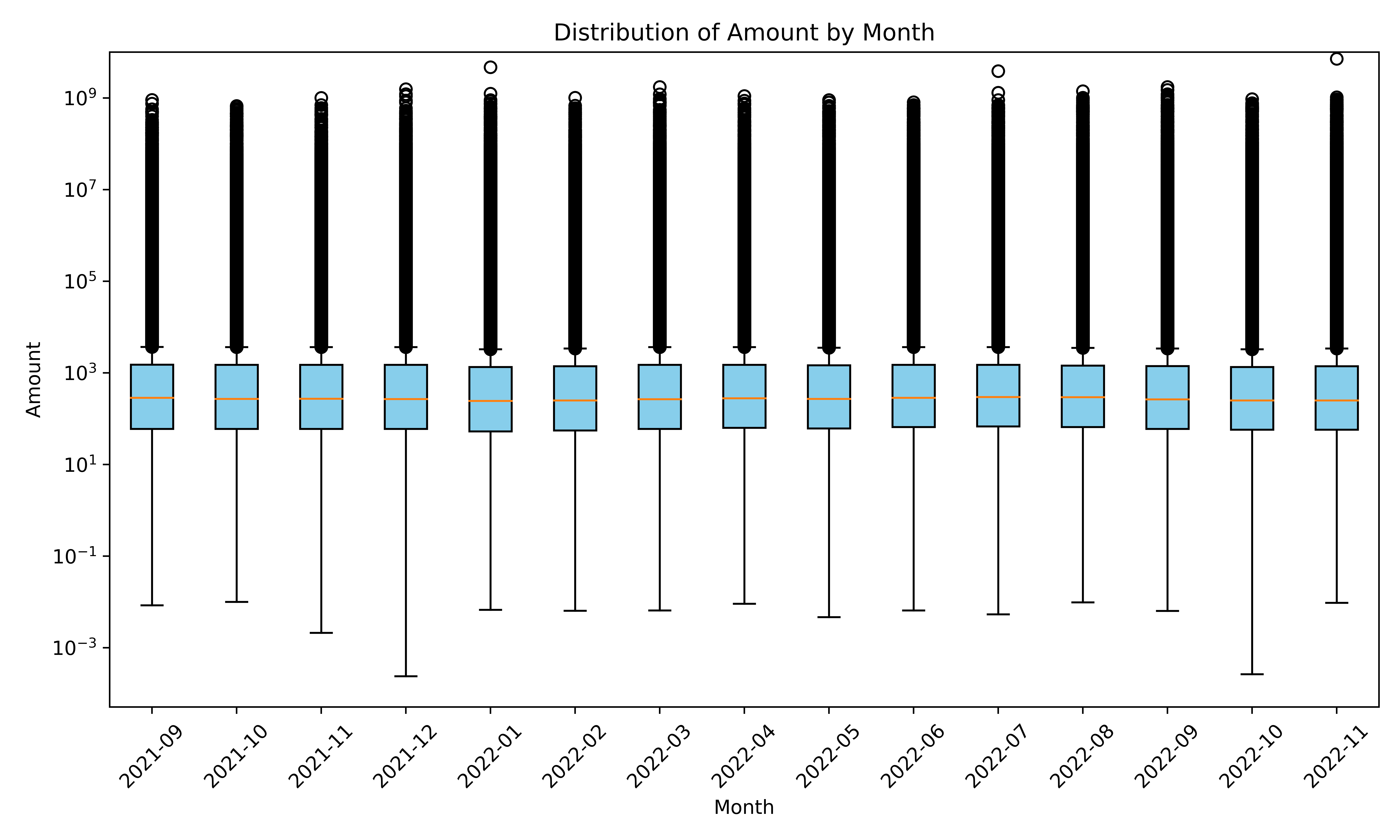} 
\caption{Distribution of the variable \textit{amount} by month.}
\label{fig:amount_distr}
\end{figure}

In this section, we provide quantitative information about the dataset used for our showcase experiment. As described in Sec.~\ref{subsec:dataset}, this dataset contains fully anonymized information about financial transactions. Specifically, each record contains a timestamp, information about the amount, currency and data stream of the transactions. By aggregating the transactions by month, the \textit{amount} variable is distributed according to Fig.~\ref{fig:amount_distr}. While the vast majority of transactions is below 1,000 euros, a considerable number of outliers can reach up to $10^9$ euros.

\begin{figure}[ht!]
\centering
\includegraphics[width=0.9\textwidth]{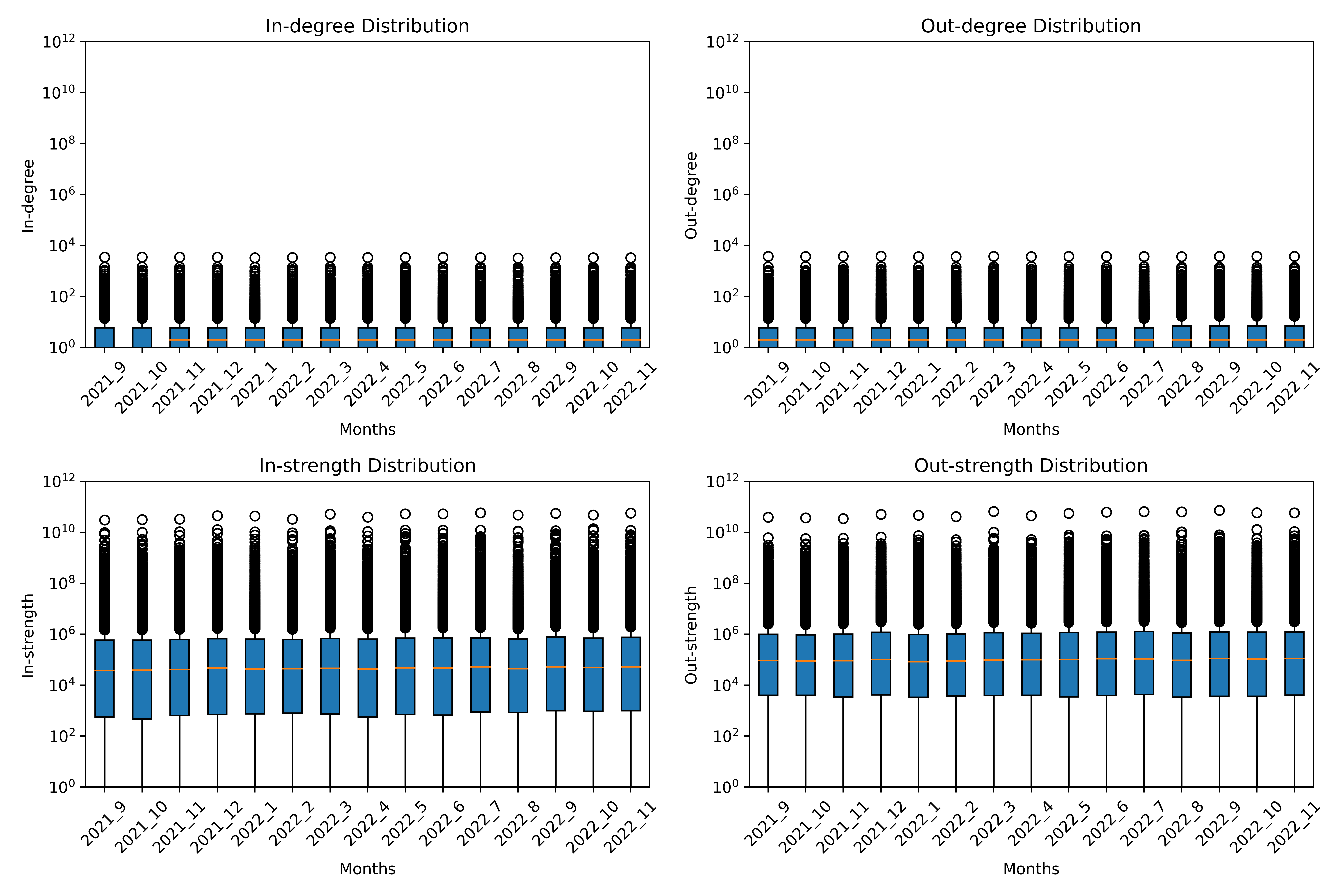} 
\caption{Distribution of indegree and outdegree, both weighted and unweighted, across the months in $G^{\mbox{BIC}}$.}
\label{fig:deg_distr}
\end{figure}

Furthermore, by aggregating the transactions by BIC, we can build the directed graphs $G_{\mbox{Sep}2021}^{\mbox{BIC}}, G_{\mbox{Oct}2021}^{\mbox{BIC}}, \ldots, G_{\mbox{Nov}2022}^{\mbox{BIC}}$. In Fig.~\ref{fig:deg_distr} we can observe the distributions of the indegree, outdegree, instrength and outstrength across the months.

\begin{figure}[ht!]
\centering
\includegraphics[width=0.9\textwidth]{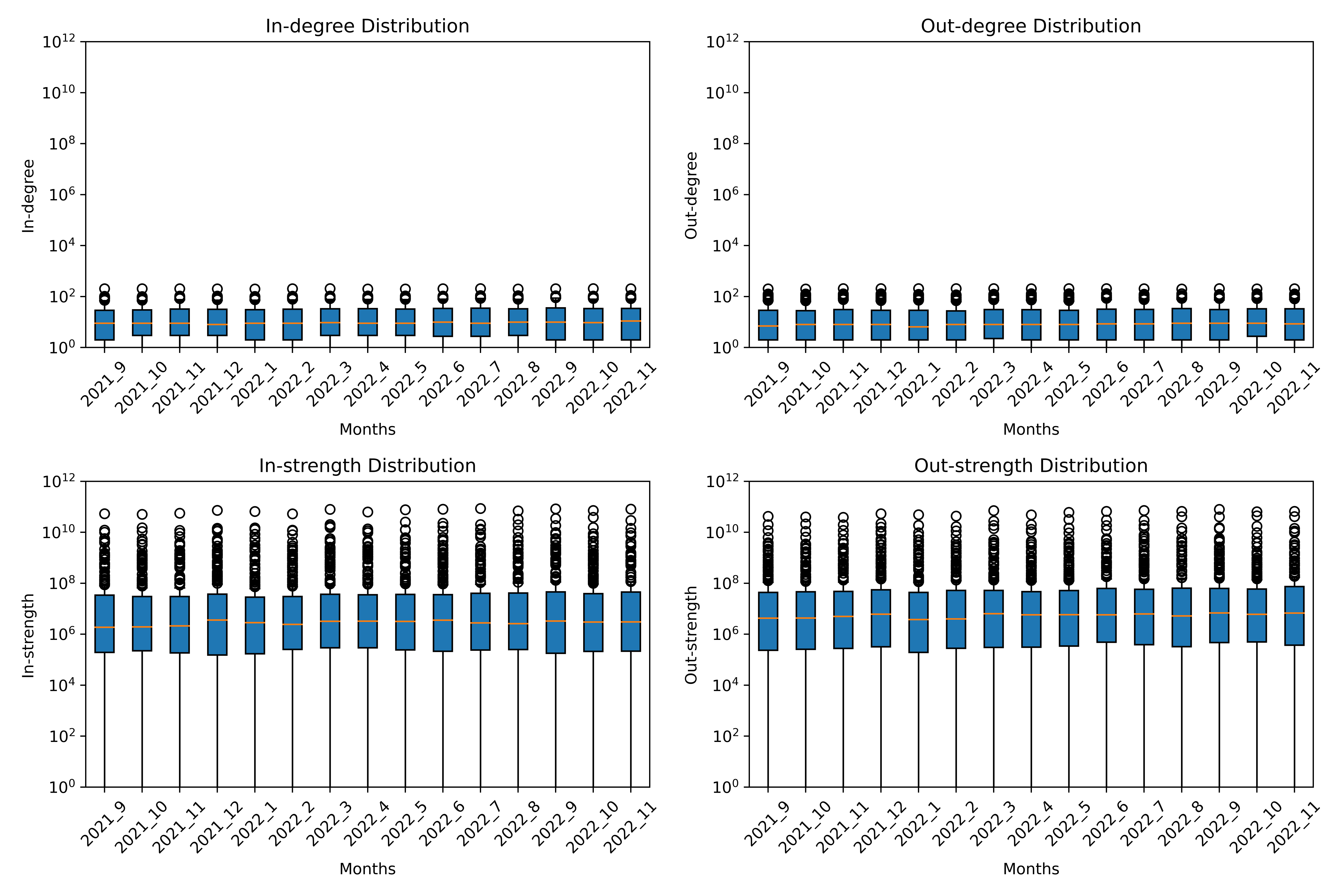} 
\caption{Distribution of indegree and outdegree, both weighted and unweighted, across the months in $G^{\mbox{Country}}$.}
\label{fig:deg_distr_ctry}
\end{figure}

\begin{figure}[h!]
\centering
\includegraphics[width=0.9\textwidth]{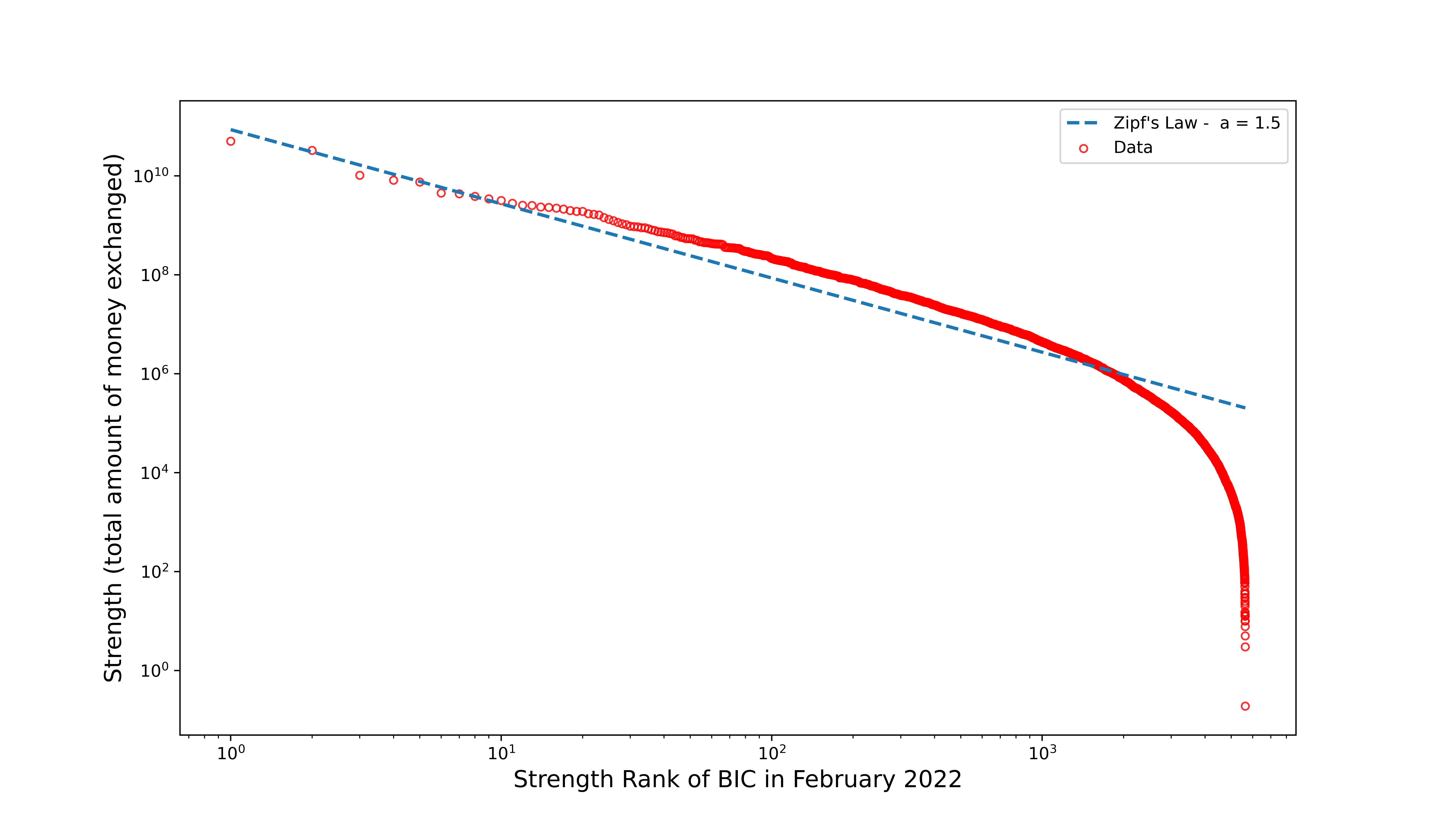}
\caption{Amount of money moved by a node against its position on the Strength ranking, fitted to a Zipf's law with shape parameter $\alpha = 1.5$.}
\label{fig:zipf_strength}
\end{figure}

In Fig.~\ref{fig:zipf_strength} we can observe the relationship between the amount of money moved by each node and its position in the rankings based on the strength. This relationship is fitted to a Zipf's Law with shape parameter $\alpha=1.5$ in a log-log space. 

\section{Comparison with machine learning based anomaly detection methods}\label{app:ML}

The strategy of selecting the top-N nodes according to the different centrality metrics can be compared to other methods of anomaly detection based on machine learning techniques. As an example, we apply two popular techniques, OneClassSVM and IsolationForest. 
We apply them as a filtering step after step 6: once the time-varying comparisons on each metric are performed, we can use these techniques on the RECs shown in Fig.~\ref{fig:step6}. Unlike what was done while applying the pipeline, we do not pick the top-N nodes that lose (\textit{\_neg}) and gain (\textit{\_pos}) positions in the rankings, but we let the ML-based algorithms choose their own outliers. This is the only difference in an otherwise straightforward process: these outliers will then go through the usual remaining steps of the pipeline, obtaining the final, stratified list.

\begin{figure}
    \centering
    \includegraphics[width = 0.6\textwidth]{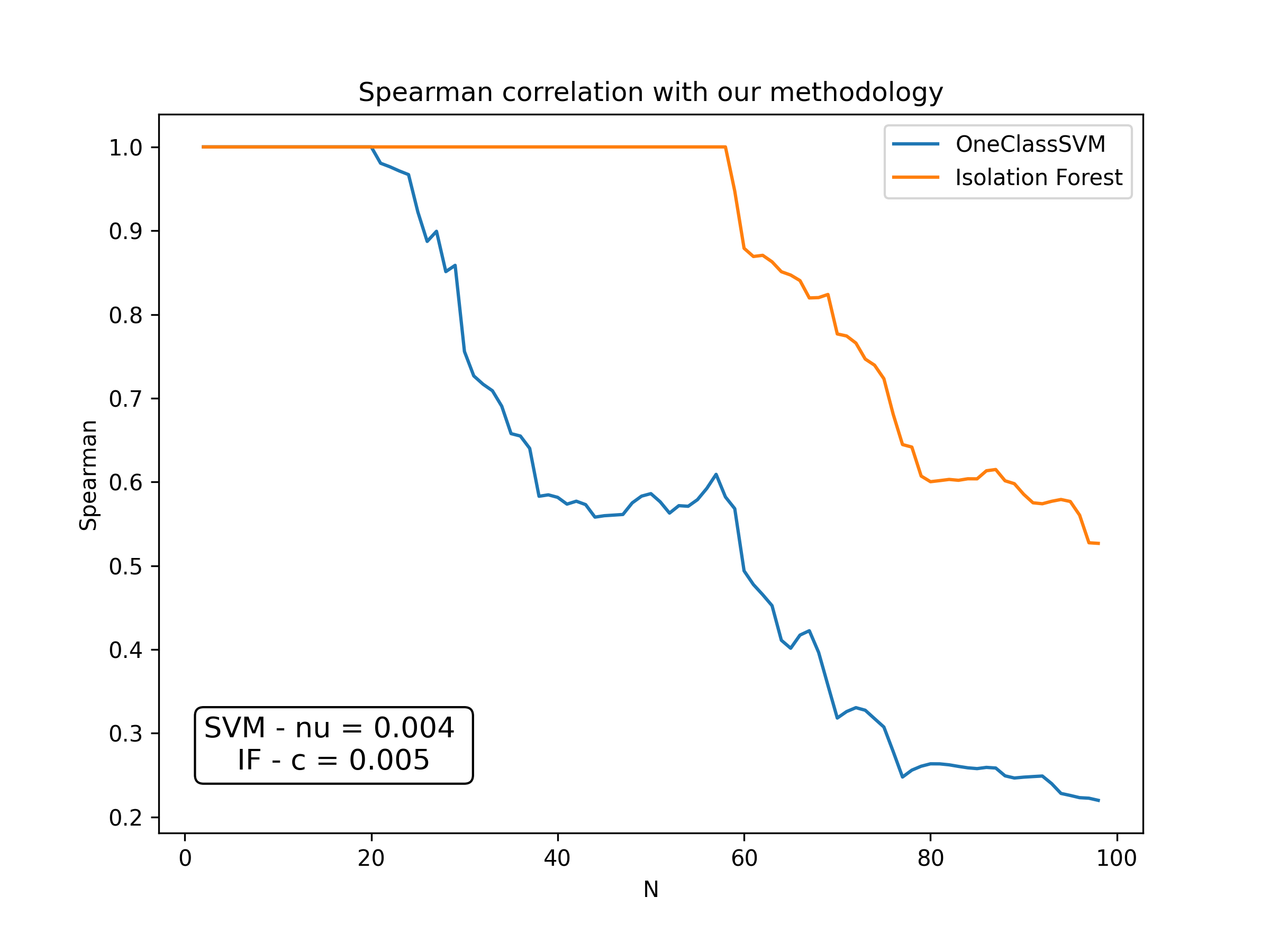}
    
    \caption{Comparison between the stratified ranking method and other machine learning based anomaly detection techniques.}
    \label{fig:nonml_comparison}
\end{figure}

 In Tab.~\ref{tab:perf_compared} we report detailed comparisons between our methodology, SVM and Isolation Forest on the individual network metrics, i.e. after their application on the RECs. The outputs obtained through OneClassSVM and Isolation Forest are no longer ranked, unlike the output of step 6 in our pipeline; therefore, we look at the performances of the individual metrics in terms of simple precision and R*. The values of $\nu$ and $c$ are chosen in order to obtain an output whose size is in the same order of magnitude of N. This task itself is not always trivial, as it appears particularly evident from the size of the output of SVM on \textit{PageRank\_neg}, which is much larger due to a cluster of data points that overlap in the same position towards the bottom of the rankings both in February and March, and are all identified as outliers by the algorithm. This non-controllability of the size of the output through SVM and IF can indeed happen depending on the inherent distribution and characteristics of the data, particularly in cases where there are clusters of overlapping points that challenge the algorithms' ability to effectively differentiate or classify them. In terms of performances, at this stage we gather that the three methodologies achieve very similar results, with Isolation Forest performing slightly better considering both the average precision and the size of the outputs across all metrics.

This similarity is confirmed by Fig.~\ref{fig:nonml_comparison}, where we focus on the comparison of the final outputs of the pipeline with and without the intermediate application of ML-based filtering methods. As we can see, the outputs have perfect overlap up to $N\simeq 380$, decreasing their correlation after that threshold as N increases. This entails that, since we are interested in the top positions of the final list of potential anomalies, there is no improvement in applying ML-based filters.

\begin{table}[]
    \centering
\resizebox{\linewidth}{!}{%
\begin{tabular}{l|cc|cc|cc|cc|cc|cc}
\toprule
{} &  avg p@1 &  r* at 1 &  avg p@5 &  r* at 5 &  avg p@10 &  r* at 10 &  avg p@30 &  r* at 30 &  avg p@50 &  r* at 50 &  avg p@100 &  r* at 100 \\
\midrule
Our method&      \textbf{1.0} &     0.05 &      1.0 &      0.1 &      0.78 &      0.14 &      0.53 &      0.29 &      0.47 &      \textbf{0.33} &       0.35 &       0.52 \\
OneClassSVM   &      1.0 &     0.05 &      1.0 &      0.1 &      0.78 &      0.14 &      \textbf{0.54} &      0.29 &      \textbf{0.54} &      0.29 &       \textbf{0.54} &       0.29 \\
Isolation Forest    &      1.0 &     0.05 &      1.0 &      0.1 &      0.78 &      0.14 &      0.53 &      0.29 &      0.47 &      \textbf{0.33} &       0.33 &       \textbf{0.57} \\
\bottomrule
\end{tabular}}
\medskip
 \caption{Performance of the final output of our method compared with OneClassSVM and IF. We notice a perfect overlap in the final outputs across the three method, suggesting that no significant improvement is brought by the application of ML-based methods.}
    \label{tab:perf_compared}
\end{table}

\section{Analysis of the IBAN subnetworks}\label{app:iban}

We can explore the network of IBANs \textit{underlying} the BICs that are flagged as anomalies as a product of the anomaly detection pipeline. Specifically, the IBAN network is built by:

\begin{enumerate}
    \item selecting a BIC;
    \item considering all the IBANs belonging to the selected BIC and the edges among them;
    \item adding and following all the edges starting from or arriving to these IBANs to include other IBANs;
    \item including all the edges between these new IBANs;
    \item adding and following all the edges starting from or arriving to these IBANs to include another set of IBANs.
\end{enumerate}

By doing so, we highlight interesting differences between the networks underlying potential anomalous and regular BICs and, most importantly, between \textit{true positive} and \textit{false positives} BICs.

As an example, we consider an analysis on some \textit{true positive} BICs, compared against \textit{false positives} and \textit{regular} BICs. Considering two different timeframes on which we are looking for anomalies (in this example, February and March 2022), we can build the IBAN networks underlying selected BICs and inspect the changes from one timeframe to the other. 
We come across two more frequent scenarios: i) the IBAN network changes dramatically in size, gaining or losing several nodes and edges, i.e., changing from a topological perspective or ii) the IBAN network does not change significantly, but inspecting it is still interesting because of the insights gained by looking at suspicious transactions.

\begin{figure*}[ht!]
\centering
    \begin{subfigure}[b]{0.49\textwidth}
        \centering
        \includegraphics[width=1.1\textwidth]{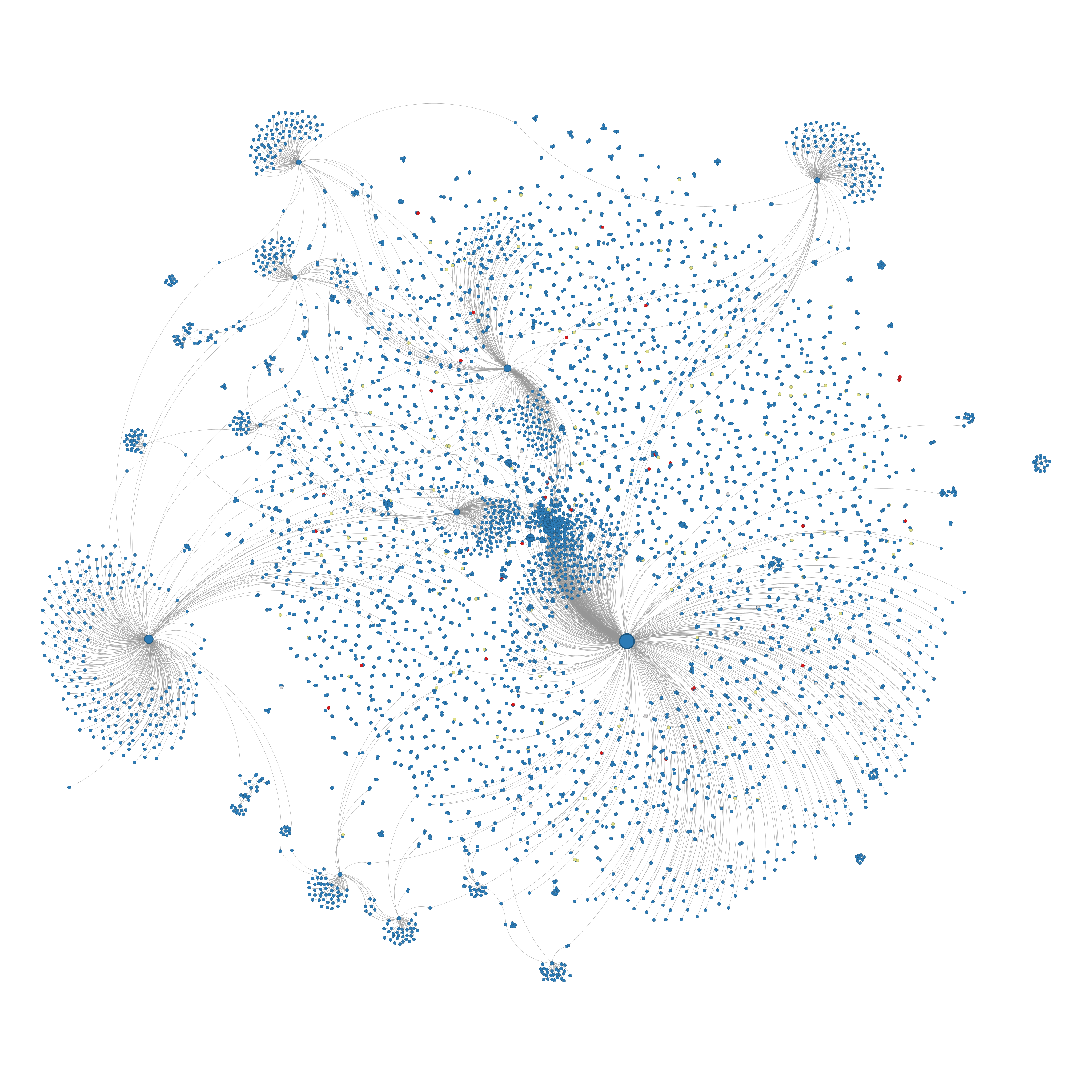}
        \caption{}
        \label{fig:jeqsit_feb}
    \end{subfigure}
    \begin{subfigure}[b]{0.49\textwidth}
        \centering
        \includegraphics[width=1.1\textwidth]{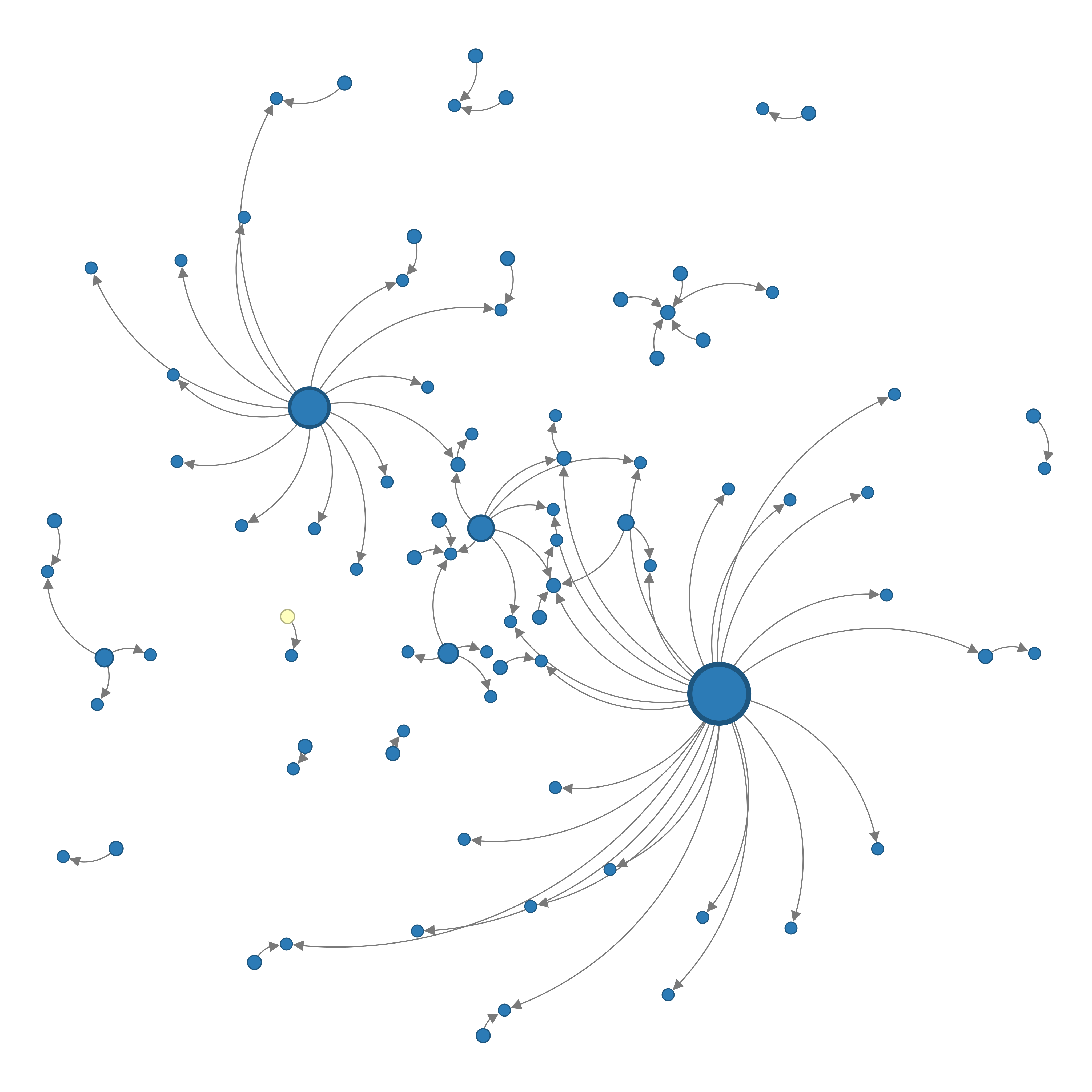}
        \caption{} 
        \label{fig:jeqsit_mar}
    \end{subfigure}
    \caption{IBAN network underlying the BIC JEQSIT21 in February (a) and March (b). We can notice the drastic drop in nodes and connections. Nodes and edges are colored according to the financial risk of their Country (green for low, yellow for medium and red for high risk).} \label{fig:jeqsit}
\end{figure*}


An example of case i) can be seen in Fig.~\ref{fig:jeqsit}. Here we build and compare the IBAN networks in February and March for BIC JEQSIT21\footnote{As in any other phase of the analysis, BICs and IBANs shown in this section are fully anonymized.}, which was marked as \textit{true positive} in the manual annotation. In this case the network goes from 6128 nodes and 4472 edges to 86 nodes and 81 edges, suggesting that it could be interesting to manually inspect the few, remaining nodes that are still involved in transactions with the anomalous BIC.

\begin{figure*}[ht!]
\centering
    \begin{subfigure}[b]{0.47\textwidth}
        \centering
        \includegraphics[width=\textwidth]{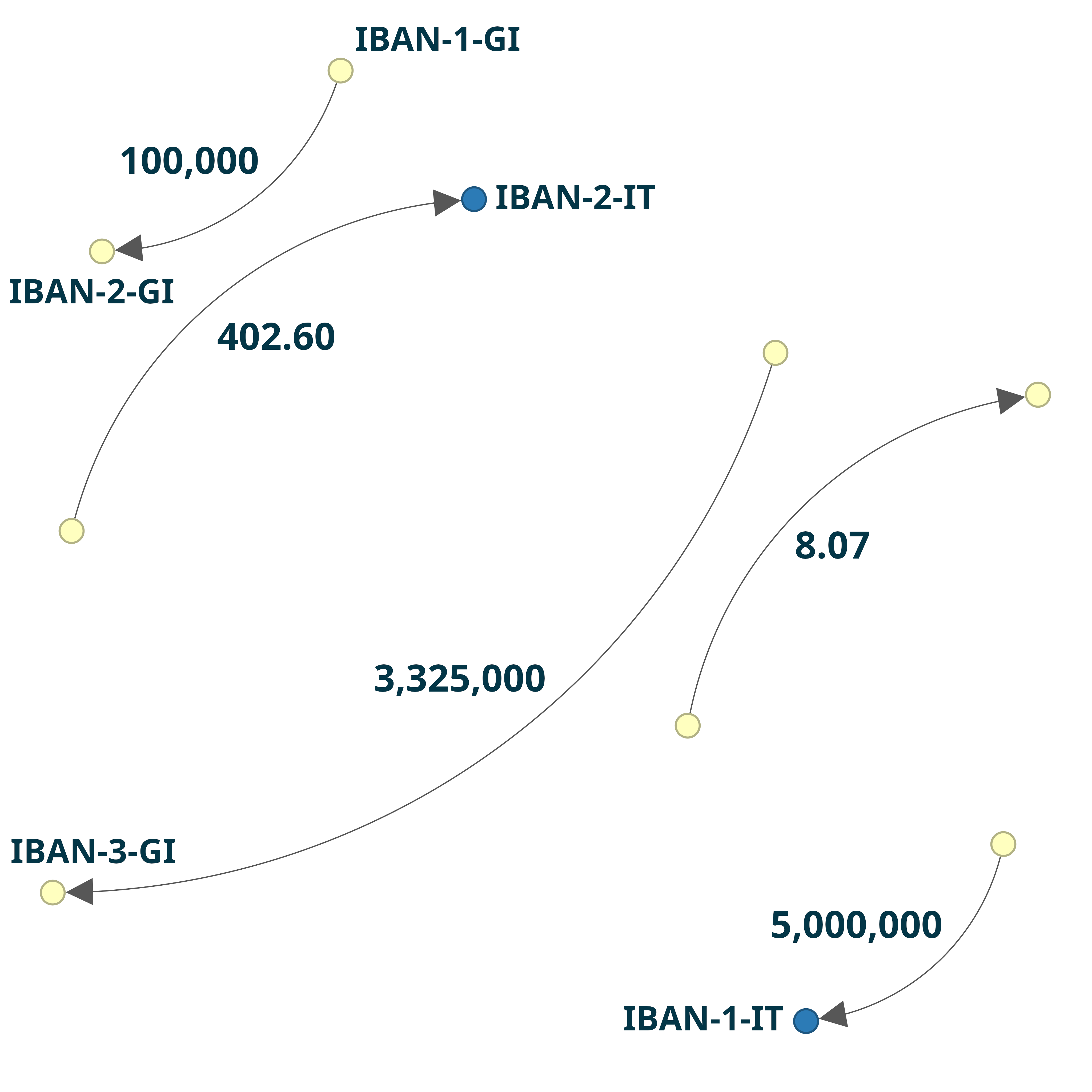}
        \caption{}
        \label{fig:jlpc_feb}
    \end{subfigure}
    \begin{subfigure}[b]{0.47\textwidth}
        \centering
        \includegraphics[width=\textwidth]{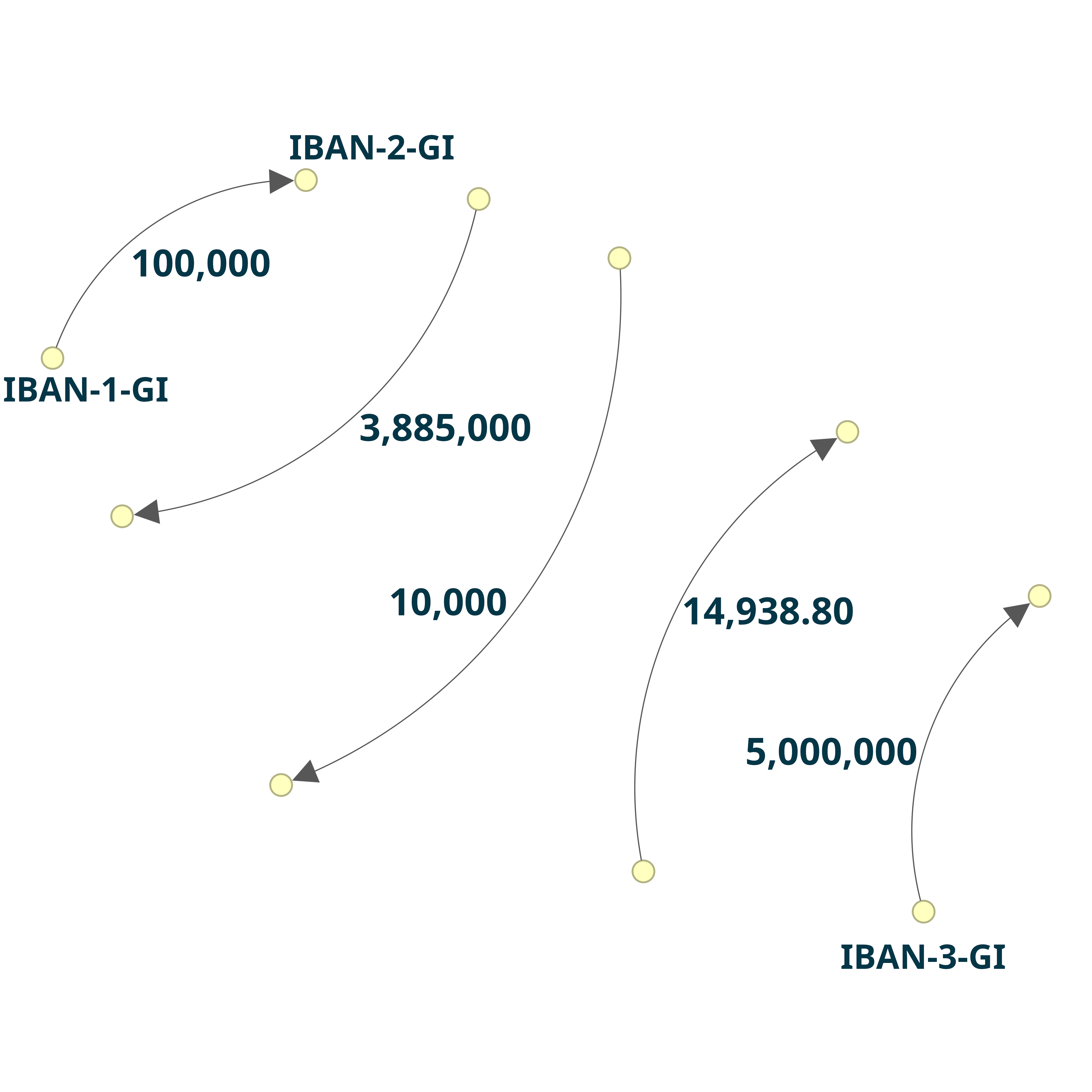}
        \caption{} 
        \label{fig:jlpc_mar}
    \end{subfigure}
    \caption{IBAN network underlying the Gibraltar BIC JLPCGISQ in February (a) and March (b). In this case the network does not change dramatically in size, but all the transactions involve countries with medium financial risk.} \label{fig:jlpc}
\end{figure*}

An example of ii) is provided in Fig.~\ref{fig:jlpc}. Here we build the IBAN networks for the Gibraltar BIC JLPCGISQ, again in February and March 2022.  In this case the network does not change dramatically in size, but all the transactions involve countries with medium financial risk, making it worth to be inspected.

\bibliographystyle{abbrv-doi}

\end{document}